\documentclass[twocolumn,epsfig, graphics,floatfix, nofootinbib]{revtex4-1}

\usepackage{amsmath,amsfonts,amssymb,graphics,graphicx,epsfig,color,times,bbm}
\usepackage{mathtools}
\usepackage{amsthm}
\usepackage{psfrag}
\usepackage{braket}
\usepackage[caption=false]{subfig}
\usepackage{graphicx}
\usepackage{color,soul}
\usepackage[dvipsnames]{xcolor}
\AtBeginDocument{\usepackage{booktabs}}
\usepackage[hidelinks]{hyperref}
\hypersetup{colorlinks=false}
\usepackage[normalem]{ulem}
\usepackage{enumitem}
\usepackage{array}

\usepackage{lipsum}

\newcommand{\zz}{\mathbbm{Z}}

\newcommand{\outcomes}{\mathcal{O}}

\newcommand{\Gsch}{G_{\mathrm{sch}}}
\newcommand{\Glog}{G_{\mathrm{log}}}
\newcommand{\val}{v}
\newcommand{\powerset}{\mathcal{P}}
\newcommand{\lspan}{\mathsf{span}}
\newcommand{\checkgen}{\Sigma}
\newcommand{\checkgrp}{\langle\Sigma\rangle}
\newcommand{\errorgrp}{\langle \mathcal{E} \rangle}  
\newcommand{\errorgen}{\mathcal{E}} 
\newcommand{\membranes}{\mathcal{M}} 

\newcommand{\dectasks}{\mathcal{T}} 

\usepackage[framemethod=tikz]{mdframed}
\usepackage{hyperref}

\definecolor{info_bgcol}{RGB}{220,237,246}
\definecolor{info_textcol}{RGB}{58,100,126}
\definecolor{info_linecol}{RGB}{220,237,246}

\mdfdefinestyle{box_style}{
  skipabove=.7\baselineskip,
  skipbelow=.7\baselineskip,
  innertopmargin=.65\baselineskip,
  innerbottommargin=.65\baselineskip,
  innerleftmargin=.5\baselineskip,
  innerrightmargin=.5\baselineskip,
  splittopskip=1.5\baselineskip,
  splitbottomskip=\baselineskip,
  roundcorner=.3\baselineskip
}
\mdfdefinestyle{info_style}{
  style=box_style,
  backgroundcolor=info_bgcol,
  linecolor=info_linecol,
  fontcolor=info_textcol,
}

\newmdenv[style=info_style]{info}

\theoremstyle{plain}
\newtheorem{thm}{Theorem}
\theoremstyle{plain}

\theoremstyle{definition}
\newtheorem{defn}{Definition}

\theoremstyle{remark}

\usepackage{multirow}

\begin{document}

\clearpage
\newpage

\pagenumbering{arabic}

\newcommand*\correspondingauthor{\thanks{Corresponding author: \mbox{fpastawski@psiquantum.com}}}

\title{Modular decoding:\\ parallelizable real-time decoding for quantum computers}
\author{H\'ector Bomb\'in}
\author{Chris Dawson}
\author{Ye-Hua Liu}
\author{Naomi Nickerson}
\author{Fernando Pastawski}\correspondingauthor
\author{Sam Roberts}

\affiliation{PsiQuantum Corp., Palo Alto}
\date\today

\begin{abstract}
    Universal fault-tolerant quantum computation will require real-time decoding algorithms capable of quickly extracting logical outcomes from the stream of data generated by noisy quantum hardware.
    We propose \textit{modular decoding}, an approach capable of addressing this challenge with minimal additional communication and without sacrificing decoding accuracy.
    We introduce the {\it edge-vertex} decomposition, a concrete instance of modular decoding for lattice-surgery style fault-tolerant blocks which is remarkably effective.
    This decomposition of the global decoding problem into sub-tasks mirrors the logical-block-network structure of a fault-tolerant quantum circuit.
    We identify the {\it buffering condition} as a key requirement controlling decoder quality; 
    it demands a sufficiently large separation (buffer) between a correction committed by a decoding sub-task and the data unavailable to it.
    We prove that the {\it fault distance} of the protocol is preserved if the buffering condition is satisfied.
    Finally, we implement edge-vertex modular decoding and apply it on a variety of quantum circuits, including the Clifford component of the 15-to-1 magic-state distillation protocol.
    Monte Carlo simulations on a range of buffer sizes provide quantitative evidence that buffers are both necessary and sufficient to guarantee decoder accuracy.
    Our results show that modular decoding meets all the practical requirements necessary to support real-world fault-tolerant quantum computers.
\end{abstract}

\maketitle

\section{Overview of results}\label{sec:Overview of results}

In this article, we present and analyze \textbf{modular decoding}, a general method for decomposing decoding problems into smaller decoding sub-tasks.
This decomposition allows meeting the practical requirements of a fault-tolerant quantum computer.
Firstly, modular decoding allows concurrent execution by construction, which allows side-stepping throughput limitations associated to any single processing core executing the decoding algorithm.
More importantly, modular decoding is a real-time (live) approach to decoding, providing intermediate logical outcomes with minimal latency.
This is an essential, yet often overlooked, requirement to enable universal fault-tolerant quantum computation.

We identify \textbf{buffering} as the key technique enabling modular decoding to produce high quality decoding results.
The \textit{buffering condition} requires the input to each decoding task to include a \textit{buffer} of additional outcome data in directions where previous decoding tasks have not set fixed boundary condition.
The buffer for each decoding task can be determined by a graph algorithm which works for arbitrary modular decoding decompositions.

We prove a rigorous \textbf{soundness theorem for modular decoding}, which guarantees that decoding on a modular decomposition can be as effective at catching errors as an \textit{offline decoder} accessing the entire decoding problem.
The only assumptions are that each decoding sub-task is solved using a sound decoder, and provided a sufficiently large buffer of outcome data (termed the \textit{buffering condition}).

We introduce \textbf{edge-vertex decoding}, a concrete instance of \textit{modular decoding} well suited to lattice-surgery style fault tolerance.
This approach follows the quantum circuit structure distinguishing two kinds of decoding sub-tasks.
Each corresponds to either an elementary logical block (\textit{vertex task}) or to a connection among neighboring elementary blocks (\textit{edge task}).
All edge decoding tasks can be solved in parallel and only require a buffer of outcome data.
Vertex decoding tasks require boundary condition data from neighboring edge tasks but can otherwise be solved independently and in parallel.
By design, the use of buffers guarantees maintaining decoding quality (soundness), whereas the minimal size of decoding sub-tasks and their data dependencies reduces logical outcome latency. 

We \textbf{implement} edge-vertex decoding decompositions \textbf{and numerically benchmark} the decoding quality against offline decoding via Monte Carlo simulations. 
We find evidence that a buffer width commensurate with the protocol distance is both necessary and sufficient to maintain the same decoding performance as offline decoding.
Our numerical simulations, which include large-scale \textbf{15 to 1 magic state distillation}, access logical block networks significantly more complex than current literature.

\subsection{Outline and readers guide}

The article is organized as follows.
\textbf{Sec.~\ref{sec:Background and motivation}} provides general background on quantum computing, fault-tolerance and decoding.
It also motivates the need for modular decoding using unitary gate synthesis as an example, and present connections with previous work.
It can be safely skipped by an expert reader who wishes to go directly to more technical material.
\textbf{Sec.~\ref{sec:Preliminaries}} establishes technical notation and concepts relevant to decoding which will allow precise formulation of algorithms and theorems in later sections.
In \textbf{Sec.~\ref{sec:MDmethods}}, we introduce the modular decoding problem, which involves decomposing the global problem into smaller, decoding sub-tasks such that their results can be straightforwardly combined into a global correction.
We show naive modular decoding effectively decreases the fault distance, and show how buffering can be used to overcome this. 
\textbf{Sec.~\ref{sec:soundness-proof}}, we prove that buffering is sufficient to maintain decoder soundness---that errors with weight up to half the fault distance can still be corrected, given a sufficient buffer.
The proof may be skipped by readers who are not mathematically inclined.
\textbf{Sec.~\ref{sec:scheduling}} we introduce methods to schedule the various modular decoding sub-tasks, and introduce the edge-vertex decoder which has very low reaction time. 
\textbf{Sec.~\ref{sec:simulations}} presents extensive simulation data for edge-vertex modular decoding, including complex logical block networks in the fault-tolerant $15$-to-$1$ magic-state distillation protocol. 
We conclude in \textbf{Sec.~\ref{sec:Conclusions_and_outlook}}, which summarizes our work and discusses directions for further research.

\section{Background and motivation}\label{sec:Background and motivation}

Universal quantum computers can efficiently solve problems which are otherwise intractable for their classical counterparts.
Noise, imperfections and errors have thus far hindered scaling up from the realm of intermediate scale quantum devices (NISQ)~\cite{Preskill2018} to a regime of useful large scale quantum computers.
Fault tolerance (FT) prescribes how to overcome this;
scale up the number of quantum components while maintaining individual error rates per physical operation {\it below threshold}.

Quantum computers seek to answer otherwise inaccessible computational problems.
In fault-tolerant quantum computing, these answers are encoded in physical measurement outcomes
and can only be reliably extracted by compensating for the diagnosed noise.
Thus the {\it logical outcome} information needs to be {\it decoded} from vast amounts of classical data. 
The {\it decoder subsystem} is responsible for this task and may be implemented by any combination of software, firmware or hardware adequately processing the classical data.

A decoder is good if it produces reliable logical outcomes.
This allows comparing decoders operating under the same fault-tolerant protocol and physical error model.
A better decoder will boast consistently lower {\it logical error rates} (LER). 

Modeling the decoder as a monolithic function or algorithm implicitly assumes that all physical measurement data can simultaneously be made available as the decoder's input.
This situation is often referred to as {\it offline decoding}, as it is compatible with storing all the \textit{physical} outcome data produced during of a fault-tolerant quantum computation and decoding \textit{logical} outcomes from it at a later time (i.e., offline).
Most academic literature assumes this offline decoding idealization; this simplifies decoder implementation and allows comparing LER and threshold performance among decoders.

However, decoded outcomes need to become available throughout the computation.
The decoding process must quickly provide partial results on the logical outcomes in terms of available physical outcomes.
In general, the entirety of the quantum circuit is not even defined as subsequent quantum gates may be chosen depending on logical outcomes obtained.
This means that, in practice, decoding can not be modeled as a monolithic offline process but rather \textit{online} as a \textit{streaming} transformation of physical outcomes into logical outcomes.

The time taken from the moment relevant physical measurement outcomes become available to the time the decoded logical outcomes can be used to control further logic is called the {\it reaction time}.
The reaction time should be kept short, as it can potentially limit the rate of logic gates (such as $T$-gates), as illustrated by the example of arbitrary angle rotations (see Fig.~\ref{fig:SR_rotation1}).
Contributions to the reaction time come from the classical processing time required for (partial) decoding as well as communication latency.

If the decoder can not keep up with the outcome data rate (also called throughput), the quantum computation may be forced to idle (i.e., perform identity gates) until a conditional follow-up operation can be determined, and in the process generate more syndrome data to be processed.
This approach leads to a {\it decoding back-log}~\cite{dennis2002topological, terhal2015quantum} growing exponentially with the depth of the logical adaptivity and the {\it reaction-time} growing exponentially as the quantum circuit progresses.
Thus, the decoders processing throughput must fully cover the outcome data throughput to keep the reaction-time constant (and hopefully small).

Which outcomes are considered relevant for decoding a logical outcome is a critical choice which impacts both the reaction time and logical error rate.
Our main contribution is to provide a \textit{modular decoding algorithm}, wherein sensible choices for relevant inputs can lead to logical error rates which are close to optimal and a uniform reaction time which is independent of the quantum algorithm being executed.

The overall decoder \textit{throughput} (i.e. the speed at which data is processed) can be increased by decomposing the global decoding problem into independent \textit{decoding sub-tasks} and distributing these among multiple \textit{decoder units} which operate in parallel.
However, the decoding tasks can not be made fully independent of each other, as their joint outcomes must be consistent with the observed syndrome.
In this article, we show that it is possible to minimize the data dependency among these decoding tasks in such a way that a reaction-time independent of the quantum circuit size can be achieved.
The joint throughput of the decoder units only needs to exceed the production rate of physical outcome data by a constant factor as some outcome data may be processed by multiple decoder units.

\subsection{Motivating example: intermediate logical outcomes in gate synthesis} 

\begin{figure*}[ht!]
\includegraphics[width=0.99\linewidth]{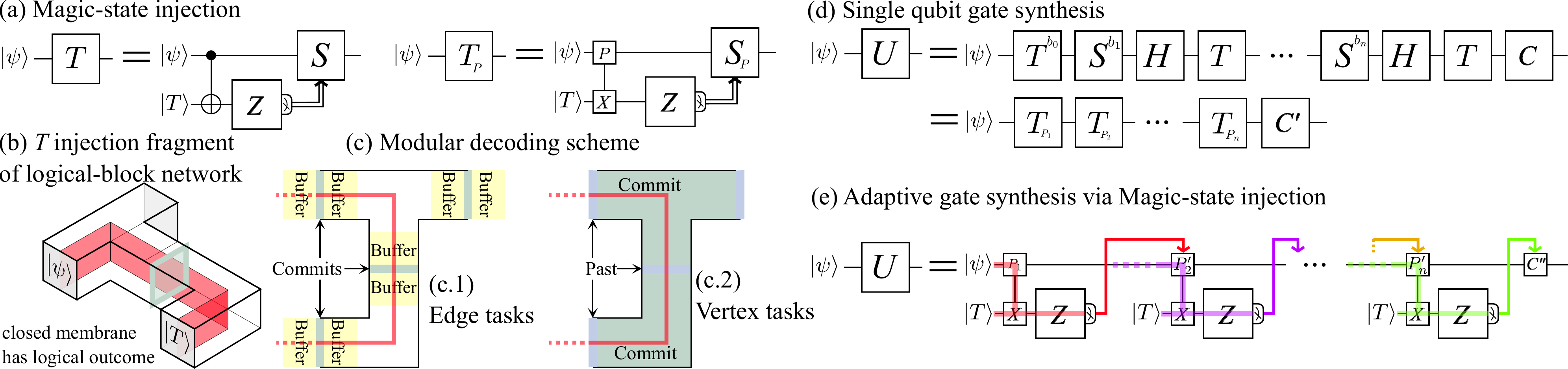}
\caption{
\textbf{(a) (left)} A $T$ gate is applied via magic state injection and requires a Clifford correction $S$ conditioned on the $Z$ measurement outcome on the second qubit.
This measurement outcome requires the Pauli frame for $Z\otimes Z$ on the inputs.
\textbf{(a) (right)} The protocol can be generalized to realize a generalized $T$ gate $T_P := \exp{\left(i \pi/8 P\right)}$
by applying a generalized CNOT.
\textbf{(b)} An isometric view of a 3D logical-block network fragment corresponding to magic state injection of (a) for a $T=T_Z$ gate. 
The outcome dependent $S$ gate correction is not included as it can be efficiently tracked classically.
The logical membrane highlighted in red closes to produce a logical outcome (the $Z$ measurement outcome of a).
The interface between two logical-blocks is highlighted in green.
Both membrane and interface are bookkeeping tools, neither having a physical signature.
\textbf{(c)} A 2D \textit{top-view} schematic for the edge-vertex modular decoding decomposition for the fragment of logical-block network presented in (b).
The \textit{edge-vertex} modular decoding decomposition first solves edge decoding tasks (c.1) using syndrome data from in neighbouring buffer regions (yellow).
After edge decoding tasks are complete, vertex decoding tasks (c.2) can be solved independently, each task having boundary conditions set by neighbouring edge tasks. 
This scheme has very fast reaction times, as there is no significant chain of dependencies between decoding tasks.
\textbf{(d)} In gate synthesis, arbitrary single qubit unitaries may be approximated by a sequence of $T$ gates interspersed  with $\{SH, H \}$ Clifford gates and a final Clifford gate $C$ or alternatively as a sequence of generalized $T$ gates.
\textbf{(e)} When $T$ gates are performed via magic state injection, each logical outcome $b_j$ obtained during injection may require an additional $S_j$ correction.
These logical outcomes correspond to partially highlighted logical membranes represented in more detail in (b) or (c).
Each of these membranes continue and branch into the past but can be efficiently summarized by the logical Pauli frame which tracks their cumulative value.
As an alternative to physically implementing $S$ corrections, these may be incorporated in a Clifford frame, modifying which generalized $T$ gate is to be performed next.
Specifically, each $P'_k$ is given by $P'_k := C_k P_k C_k^{\dagger}$, with 
$C_k := S^{b_1}_{P_1}\ldots S^{b_{k-1}}_{P_{k-1}}$.
Similarly $C'':= C_n C' C_n^\dagger$, which need not be physically implemented.
Colored arrows illustrate when and where the logical outcomes $b_j$ are needed to condition further logical blocks.
}
\label{fig:SR_rotation1}
\end{figure*}

To motivate the need for low-latency decoders (i.e., low reaction time), we consider the problem of \textit{gate synthesis} in surface code fault-tolerance.
This refers to approximating arbitrary single-qubit gates as products of $T$ gates ($\pi/8$-rotations in the $Z$ basis) and Clifford operations.
This decomposition is representative and of widespread practical relevance.
As we will see, a natural implementation of this sub-routine involves adaptive sequences of operations wherein each operation is conditioned on a logical measurement outcome from the immediately preceding one.
This motivates the need for a decoding approach which can provide logical measurement outcomes as soon as possible.

The Matsumoto and Amano normal form \cite{Matsumoto2008, Giles2013} is presented in the first line of Fig.~\ref{fig:SR_rotation1} (d) and can be used to express arbitrary products of Clifford and $T$ gates in a way that minimizes the number of $T$ gates.
An operator $H S^{b_j}$, with $b_j \in \{0,1\}$ is applied between otherwise consecutive $T$ gates.
In particular, efficient (in number of $T$ gates) approximations of single qubit unitaries such as those used for gate synthesis can be presented in this form.
Selinger and Ross provide an efficient algorithm to find the exponents $b_j$ best approximating rotations around the $Z$ axis~\cite{ross2016optimal}. 

As an alternative to the Matsumoto and Amano form, the intermediate Clifford operators can be avoided by appropriately conjugating each of the $T$ gates involved in the sequence.
This results in a sequence of $n$ generalized $T$ gates $T_P := \exp{\left(i \pi/8 P\right)}$, followed by a single Clifford operator $C'$ as shown in the second line of Fig.~\ref{fig:SR_rotation1} (d).
Valid sequences $C' T_{P_n}\ldots T_{P_2} T_{P_1}$ are characterized by two conditions: i) $P_i \in \{X, Y, Z\}$ (i.e., no minus signs on Paulis) and ii) consecutive gates $T_{P_i}, T_{P_i+1}$ are distinct ($P_i \neq P_{i+1}$).

Neither $T$ gates nor generalized $T_P$ gates can be achieved exclusively with lattice surgery.
However, each $T$-gate can be realized by injecting a distilled magic-state $\ket{T}$, as shown in Fig.~\ref{fig:SR_rotation1} (a)(left), whereby a measurement is performed which teleports the $T$ gate into the data qubit up to a conditional $S$ gate correction.
A generalized $T_P$ gate (instead of $T\equiv T_Z$) can also be realized in a similar manner as shown in Fig.~\ref{fig:SR_rotation1} (a)(right).
To do so, the CNOT and $S$ unitary operations in the injection protocol must be replaced by  Clifford $C$ conjugated versions (where $C$ acts on the first qubit such that $P = CZC^\dagger$).
In other words, instead of a CNOT operator, a generalized C$_P$NOT $=(II + PI + IX -PX)/2$ is applied and the correction $S_P := \exp{\left(i \pi/4 P\right)}$ (instead of $S$) is required conditional on the measurement outcome.

If generalized $T_P$ gates are performed via magic state injection, then each one will involve the measurement of a logical outcome and the application of a $S_P$ correction conditioned on the outcome of said measurement.
Logical outcomes will also be needed immediately if $S_P$ corrections are tracked as a Clifford frame (i.e. conjugate all forthcoming operations) rather than performing them explicitly.
This is illustrated in Fig.~\ref{fig:SR_rotation1}(e), where each logical outcome condition the choice of the next generalized $T_P$ gate.
The auto-corrected $\pi/8$ gates introduced in \cite{litinski2019game} are a resource efficient alternative which reaps the most benefit from the short reaction times which can be achieved via modular decoding.

For concreteness, we illustrate fault-tolerant quantum computations using \textit{surface codes} and \textit{lattice surgery}~\cite{horsman2012surface,litinski2019game}, where quantum algorithms are represented by 3D space-time networks of logical blocks~\cite{bombin2021logical}, each element of which is a 3D block representing a FT quantum instrument. 
Such blocks have input/output ports which can be composed across to represent quantum circuit topology. 

Logical measurement outcomes are obtained by combining measurement outcomes of specific physical measurements.
We refer here to the set of physical measurement outcomes making up a logical measurement as a \textit{logical membrane} due to the geometric shape of its support. 
For example, the logical outcome of measuring $Z\otimes Z$ on $\ket{\psi}\otimes \ket{T}$, is associated with the schematically depicted (red) logical membrane in Fig.~\ref{fig:SR_rotation1} (b).
This is equivalent to the single qubit $Z$ a measurement on the second qubit which occurs after a CNOT operator in Fig..~\ref{fig:SR_rotation1} (a).
The membrane in Fig. ~\ref{fig:SR_rotation1} (b) continues towards the past and depends on the history of the two qubits.
For our purpose, we may imagine that this history is summarized by two bits which accumulate the parities of the two membrane continuations.

We can decompose the lattice surgery into smaller constituent blocks which extend $\approx d$ in each direction. 
This is the decomposition is referred to as a \textit{logical-block network} decomposition.
The connectivity structure of this network is succinctly conveyed by a graph,
with elementary blocks and their connected ports interpreted as vertices and edges respectively.

The two connected logical blocks in Fig.~\ref{fig:SR_rotation1} (b) are only a fragment of the larger network for a full quantum algorithm or sub-routine.
In order to provide a logical outcome, information from a set of logical blocks supporting the corresponding membrane will need to be decoded.
Fig.~\ref{fig:SR_rotation1} (e) partially illustrates how the support could look in terms of elementary circuit components for the example of gate synthesis using magic state injection.
Fig.~\ref{fig:SR_rotation1} (e) also emphasizes how quickly logical outcomes are needed to condition subsequent quantum operations.
Quantum sub-routines will, in general, require \textit{classical control}~\cite{PERDRIX_2006}, i.e. using available logical outcomes to control subsequent quantum logic.

The example in Fig.~\ref{fig:SR_rotation1} (e) and other quantum algorithms and subroutines involve decoding on large logical block networks with stringent requirements to enable classical control.
Our modular-decoding framework allows decomposing the decoding problem into sub-tasks with the goal of enabling quick extraction of logical outcomes (i.e. fast reaction time) while retaining the low logical error rates (LER) of off-line decoders.
Edge-vertex decoding, illustrated in  Fig.~\ref{fig:SR_rotation1} (c), is a concrete modular decoding decomposition which achieves these goals for general lattice surgery style circuits.

\subsection{Prior work}

The back-log issue was highlighted in Ref.~\cite{dennis2002topological}, whereby decoding the entire history of the computation is impractical. 
They show it is possible to divide the global decoding problem into sub-decoding problems in order to get up-to-date information about the correction, even when syndrome histories involve many logical qubits. 
In Ref.~\cite{fowler2013minimum}, the use of parallelization is proposed to decrease the decoding time complexity. 
In Ref.~\cite{bartolucci2021fusion}, proposed using multiple decoders in parallel with message passing to reduce the reaction time.

More recently, in Refs.~\cite{tan2022scalable,skoric2022parallel}, a parallel approach to decoding is proposed and numerically benchmarked, showing that decoding can be parallelized without significant impact on accuracy (LER).
We note that parallel decoders for some (non-topological) quantum LDPC codes have previously been proposed~\cite{leverrier2022parallel}.

Our work was developed independently, but we note partial overlap with the recently published results in Refs.~\cite{tan2022scalable,skoric2022parallel}.
Furthermore, this article makes several contributions beyond existing literature.
First of all, our \textit{edge-vertex} modular decomposition can be generally applied to arbitrary computations implemented through topological stabilizer fault-tolerance.
Furthermore, the statement of modular decoding is sufficiently careful to be applicable to all known flavors of topological stabilizer computation~\cite{flavoursinprep}, such as circuit based quantum computation, measurement-based quantum computation~\cite{raussendorf2007topological, bolt2016foliated, nickerson2018measurement, brown2020universal}, fusion-based quantum computation~\cite{bartolucci2021fusion} and floquet-based quantum computation~\cite{hastings2021dynamically, paetznick2022performance}.
Finally, we prove a soundness theorem for modular decoding whose hypotheses give sufficient conditions for a sound decomposition of the decoding problem.

\section{Preliminaries}\label{sec:Preliminaries}

\textbf{Classical outcomes} provide the only classical window onto the quantum evolution which takes place in the computer.
It is the combination of these outcomes and a physical model for operations and errors on them which will guide the fault-tolerant control and interpretation of the computation.
Decoding deals with raw classical data produced during a quantum computation. 
These are all the outcomes from measurements performed throughout the computation, or more generally the outcomes from quantum instruments~\cite{bombin2021logical,flavoursinprep}. 
We index the collection of physical outcomes of a computation by the set $\outcomes$.
For simplicity, and in anticipation of the stabilizer formalism, we will assume that outcomes take binary values denoted by $\val: \outcomes \rightarrow \zz_2$ (with $\zz_2 = \{0,1\}$).

\textbf{Parity checks} are constraints among the measurement outcomes reflecting the redundancy among them under ideal operations (i.e., in the absence of errors). 
In stabilizer fault tolerance, a check $\sigma$ is characterized by a sub-set of outcomes with a fixed joint parity (which we will assume to be 0 without loss of generality). 
By abuse of notation, extending the definition of $\val$ linearly over subsets of $\outcomes$, we will denote this as constraint as, $\val (\sigma) = 0$. 
A value of $\val (\sigma)$ other than $0$ indicates the presence of errors. 
We are interested in the context of stabilizer fault tolerance, where the set of all checks $\checkgrp$ has the structure of a linear space generated by a given set of check generators $\checkgen$ (i.e., $\checkgrp = \lspan(\checkgen)$).
In turn, $\checkgrp$ is a linear subspace of the power-set $\powerset(\outcomes)$ of $\outcomes$, the set of all possible subsets of $\outcomes$ which can be interpreted as the linear space $\zz_2^{\outcomes}$ (the space of functions from $\outcomes$ to $\zz_2$).
In other words, $\checkgrp$ is closed under \emph{exclusive or} (XOR) of its elements (i.e., $\sigma_1 \in \checkgrp$ and $\sigma_2 \in \checkgrp$ implies $\sigma_1 \oplus \sigma_2 \in \checkgrp$).
All the information which is relevant for the decoder to identify an error is contained in $\val\vert_{\checkgrp }$ (i.e., the restriction of $\val$ to elements in $\checkgrp$).
This information can be succinctly encoded in the resulting parities for the, possibly over-complete, generating set of checks $\checkgen$.
We will assume that the decoder has access to a convenient generating set of checks $\checkgen$ as well as their syndrome outcome $\val: \checkgen \rightarrow \zz_2$ as they become available.

Although $\checkgrp$ encapsulates fundamental, decoder independent properties, the choice of $\checkgen$ has practical importance.
Check generators in $\checkgen$ impact the performance of the decoder and imprints a notion of locality which guide the decomposition of the decoding task.
Some ambiguity arises when there are multiple levels (hierarchy) of encoding involved. 
In this situation, it is possible for check generators at a higher level to be treated as logical outcomes by a lower level fault-tolerance scheme.
Magic state distillation is a practically relevant example of such a situation; there, physical injected states are unprotected by the underlying low level surface code style fault tolerance.
However, the distilled magic states are nevertheless protected by higher level distillation checks.

\textbf{Logical outcomes} are the final product of fault-tolerant quantum computation.
In the desired regime of operation, these outcomes reproduce the same value distribution as a noiseless quantum computer.
Logical outcomes $\mathcal{M}$ are derived from physical outcomes in $\outcomes$.
For stabilizer fault tolerance, logical outcomes $M \in \mathcal{M}$ take bit values.
Moreover, in the absence of errors their value $\val(M)$ is the joint parity of a corresponding subset of outcomes $M \subset \outcomes$, which gives the ideal logical outcome distribution taken as reference.
The choice of letter, $\mathcal{M}$, is intended to convey {\it membrane}, which correspond to the 2D arrangement of physical outcomes associated to logical outcomes in flavours of surface-code based fault tolerance~\cite{flavoursinprep}.

In surface-code based flavours of fault-tolerance, logical outcomes correspond to relatively closed membranes.
A membrane is relatively closed if (at some point in the computation) its support is fully contained within the set of available outcomes (i.e., it will have no support on outcomes to be generated in the future).
In general, the entirety of a logical membrane is an unwieldy object, spanning and branching through (potentially the entirety of) the computation.
The logical {\it Pauli frame} is a convenient way to tame this complexity.
The Pauli frame summarizes the parity accumulated by \textit{partial} logical membranes which may or may not, in the future, be completed into a logical outcome of $\mathcal{M}$.
In a snapshot of the computation, there would be one partial logical membrane per element of the instantaneous $n$-qubit logical Pauli group (with $n$ equal to the number of logical qubits at that instant in time); these are the bits of the Pauli frame.
This approach provides the flexibility of defining the logical circuit on the go.
When, completing a logical measurement (i.e., by closing a membrane $M$), the outcome $\val(M)$ can be obtained by taking a joint parity of the {\it historic} parity of $M$, as given by a corresponding Pauli frame element with a {\it recent} term associated to the closure of $M$.
Partition of historic and recent is arbitrary but serves a practical purpose.

For ideal noiseless operations, each bit of the Pauli frame simply accumulates the partial outcome parity for some partial membrane $M$.  
In noisy fault-tolerant computation, this parity may require adjusting in order to compensate the effect of diagnosed errors on the logical Pauli frame.

\textbf{Error model.}
In a real-world device, operations are not perfect, and so the observed outcome distribution differs from  the ideal one.
Fault-tolerant schemes can deal with such imperfections contingent on an adequately benign noise model.
Soundness proofs and numerical simulations in this and other articles assume such noise models, which are themselves an idealization.
While in this way, we gain confidence that fault-tolerance is capable of producing robust computational outcomes, the final validation will only come with an actual fault-tolerant computation.

As in most stabilizer fault-tolerance literature, we assume an error model defined in terms of a set of elementary \textit{errors} or faults $e \in \errorgen$ which are Pauli (product) operators acting on the qubits at different times throughout the computation.
This choice is motivated by the fact that for stabilizer fault-tolerance, wherein the (Pauli) measurements used to generate the parity checks collapse many coherent errors into Pauli errors, and thus the effects of a wide range of errors can be modeled in this way.
The choice of elementary errors $\errorgen$ is model specific and should aim to represent the physics and imperfections of the device(s) being modeled.

A common choice of error generators $\errorgen$, is to have a one to one relation between physical measurement outcomes $o \in \outcomes$ and each error generator $e\in \errorgen$, with each element $e$ flipping a single outcome $o$.
While this can be an adequate model certain implementations of MBQC or FBQC, it actually conflates two distinct objects and can not accurately model many other scenarios.
To comply with the format accepted by a concrete decoding algorithm (e.g., syndrome graph structure, \textit{i.i.d.} error model), the internal decoder error model may in turn be an approximation of the mathematical error model.
For simplicity, we do not distinguish them here. 

We denote by $\errorgrp$, the possible combinations of elementary errors in $\errorgen$, and will use $\epsilon \in \errorgrp$ to denote the unknown error combination occurring on the quantum system.
We further assume that the probability of different errors $\epsilon \in \errorgrp$ occurring can be described by an independent probability distribution (or possibly a low correlation distribution) on elementary error generators $e \in \errorgen$.

In a quantum instrument network (QIN), composed of stabilizer instruments (a generalization of Clifford maps including measurement)~\cite{bombin2021logical}, the effect of inserting Pauli errors between instruments is to flip a set of outcomes.
Specifically, there is a linear relation which determines which checks $\sigma \in \checkgrp$ and logical membranes $M \in \mathcal{M}$ are flipped by a given error $\epsilon \in \errorgrp$.
We denote this relation by the bi-linear map $\partial$, with $\partial \epsilon: \checkgrp \rightarrow \zz_2$, along with $\epsilon_{\mathcal{M}}: \mathcal{M} \rightarrow \zz_2$, which are themselves linear in $\epsilon\in \errorgrp$.
Thus, a specific error $\epsilon$ flips a well defined set of logical outcomes specified by the indicator function $\epsilon_{\mathcal{M}}$; the problem we are faced with is to identify which these are without knowing $\epsilon$.

\textbf{Decoding} consist of accurately inferring $\epsilon_{\mathcal{M}}$ by using the prior distribution of $\epsilon$ as well as its signature over the available syndrome information $\partial \epsilon = \val\vert_\Sigma$.
In practice, decoders approach this problem by inferring a recovery operator $\kappa$ which also belongs to $\langle \mathcal{E} \rangle$.
As such, it also has associated linear maps $\partial \kappa$ and $\kappa_{\mathcal{M}}$.
The decoder will be successful if it can identify a correction $\kappa$ such that $\kappa_{\mathcal{M}} = \epsilon_{\mathcal{M}}$.
Note that this is a much weaker requirement than $\kappa = \epsilon$, which is not necessary for successful decoding.
However, since decoders do not have access to $\epsilon_{\mathcal{M}}$ they pick a recovery operator $\kappa$ with relatively high likelihood such that $\partial \kappa = \partial \epsilon$ (i.e., the recovery operator $\kappa$ has exactly the same syndrome as the error $\epsilon$).
In other words, applying the combination of error $\epsilon$ and recovery $\kappa$ leads to all parity checks in $\Sigma$ being satisfied (i.e., the trivial syndrome).
Note however, that only $\kappa_\mathcal{M}$ and not the entirety of $\kappa$ is the relevant information produced by the decoder.
In a real-time decoding process, this allows efficiently summarizing partial decoding progress by keeping track of the effect of $\kappa$ on partial membranes with an error-corrected Pauli frame.

\begin{info}
\begin{itemize}[leftmargin=3ex]
    \item[$\epsilon$:] the physical error configuration.
    \item[$\kappa$:] the global correction chosen by the decoder.
\end{itemize}
\end{info}

\textbf{Fault distance} is an important quantity characterizing the noise resilience of a fault-tolerant protocol.
It depends only on the error generators $\mathcal{E}$ and its relationship with the logical outcomes $\mathcal{M}$ and the space of checks $\Sigma$ but is independent of the decoder.

We say that an error $\epsilon \in \langle \mathcal{E} \rangle$ is undetectable, if it has a trivial syndrome (i.e., $\partial \epsilon =0$).
The weight of an error operator $\epsilon \in \langle \mathcal{E} \rangle$, denoted by $|\epsilon|$, is the smallest number of elementary faults $e \in \mathcal{E}$ needed to express $\epsilon$ (i.e., as $\epsilon = e_1 e_2 \ldots e_{|\epsilon|}$).
The fault distance $d$ of a protocol, is the weight of the smallest non-trivial undetectable error (i.e., the smallest $|\epsilon|$ such that $\partial \epsilon = 0$ but $\epsilon_\mathcal{M} \neq 0$).
In other words, it is the weight of the smallest undetectable error giving rise to a change in a logical measurement outcome. 

Note that the weight is dependent on the error model of interest through the set of elementary error generators $\mathcal{E}$. 
If a decoder with \textit{effective fault distance} of $d$ over a fault-tolerance protocol, can {\bf detect} any error of weight up to $d-1$ and can {\bf correct} any error with weight up to $\lfloor \frac{d-1}{2} \rfloor$.
A \textit{decoder is sound} w.r.t. a fault-tolerant protocol with fault distance $d$, iff the decoder also has an effective fault distance $d$.

\begin{defn}\label{def:decsound}[{\bf Decoder soundness}]
The decoder produces a minimum weight correction $\kappa$ for any (physical) error $\epsilon$ with $|\epsilon| < d/2$, where $d$ is the fault distance of the protocol.
\end{defn}

This means that a decoder satisfying the soundness condition will correctly recover from any physical error configuration $\epsilon \in \errorgrp$ with $|\epsilon| < d/2$.
MWPM~\cite{dennis2002topological,kolmogorov2009blossom} and UF~\cite{ delfosse2017almost} are sound decoders w.r.t. any fault tolerant protocol as long as the relation $\partial$ between error generators $\errorgen$ and check generators $\checkgen$ is accurately captured by a {\it syndrome graph}.

\textbf{Syndrome graphs} are a useful data structure to represent the decoding problem whenever each elementary error $e \in \errorgen$ flips at-most two parity checks $\sigma \in \checkgen$.
This leads to a graph structure with check generators $\checkgen$ corresponding to vertices of the graph and elementary errors  $e \in \errorgen$ flipping a given pair of check generators corresponding to edges between corresponding the vertices of the graph.
One can define syndrome graphs for simple \textit{i.i.d.} Pauli and measurement error models for circuit-based quantum computation (CBQC) with the surface code~\cite{kitaev2003fault,kitaev1997quantum,dennis2002topological,raussendorf2007fault,fowler2012surface}, measurement-based quantum computation (MBQC) with topological cluster states~\cite{raussendorf2007topological,bolt2016foliated,nickerson2018measurement,brown2020universal}, or fusion-based quantum computation (FBQC) with the $6$-ring fusion network~\cite{bartolucci2021fusion,bombin2021interleaving,bombin2021logical}). 
While our numerical implementation and simulation results (Sec. \ref{sec:simulations}) make use of the syndrome graph structure, our modular decoding decomposition approach and proof (Sec. \ref{sec:soundness-proof}) are applicable more broadly to Tanner graphs with locality structure.

\textbf{Connected error clusters:} 
The check generators $\checkgen$ and error generators $\errorgen$ together with their anti-commutation relations $\partial$  define a Tanner graph.
This is a bipartite graph with each node representing either a check generator $\sigma\in \checkgen$ or an error generator $e \in \errorgen$ and edges between them used to represent the relation $\partial e(\sigma)=1$.
We will say that two error generators $e_1$ and $e_2$ are {\it directly connected} if they are distance 2 in the Tanner graph (i.e., there is at least one check generator $\sigma$ s.t. $\partial e_1(\sigma)=1$ and $\partial e_2(\sigma)=1$). 

Given an error configuration $\epsilon$ we may use this notion of {\it connectedness} to partition it into {\it connected components}.
We will call each of these components a {\it connected error cluster}.
In the case of syndrome graphs, this notion coincides with viewing $\epsilon$ as a sub-graph (subset of edges) of the syndrome graph and identifying its connected components.
The rationale behind this definition is that if \(\partial \epsilon = 0\) and \(\epsilon\) has connected components \(\epsilon_i\), then \(\partial \epsilon_i=0\). 
In other words, an error configuration $\epsilon$ is undetectable, if and only if, all of its connected components are undetectable.
For this reason, lowest weight undetectable logical errors will always consist of a single connected error cluster.

\begin{figure}
    \centering
    \includegraphics[width=0.85\linewidth]{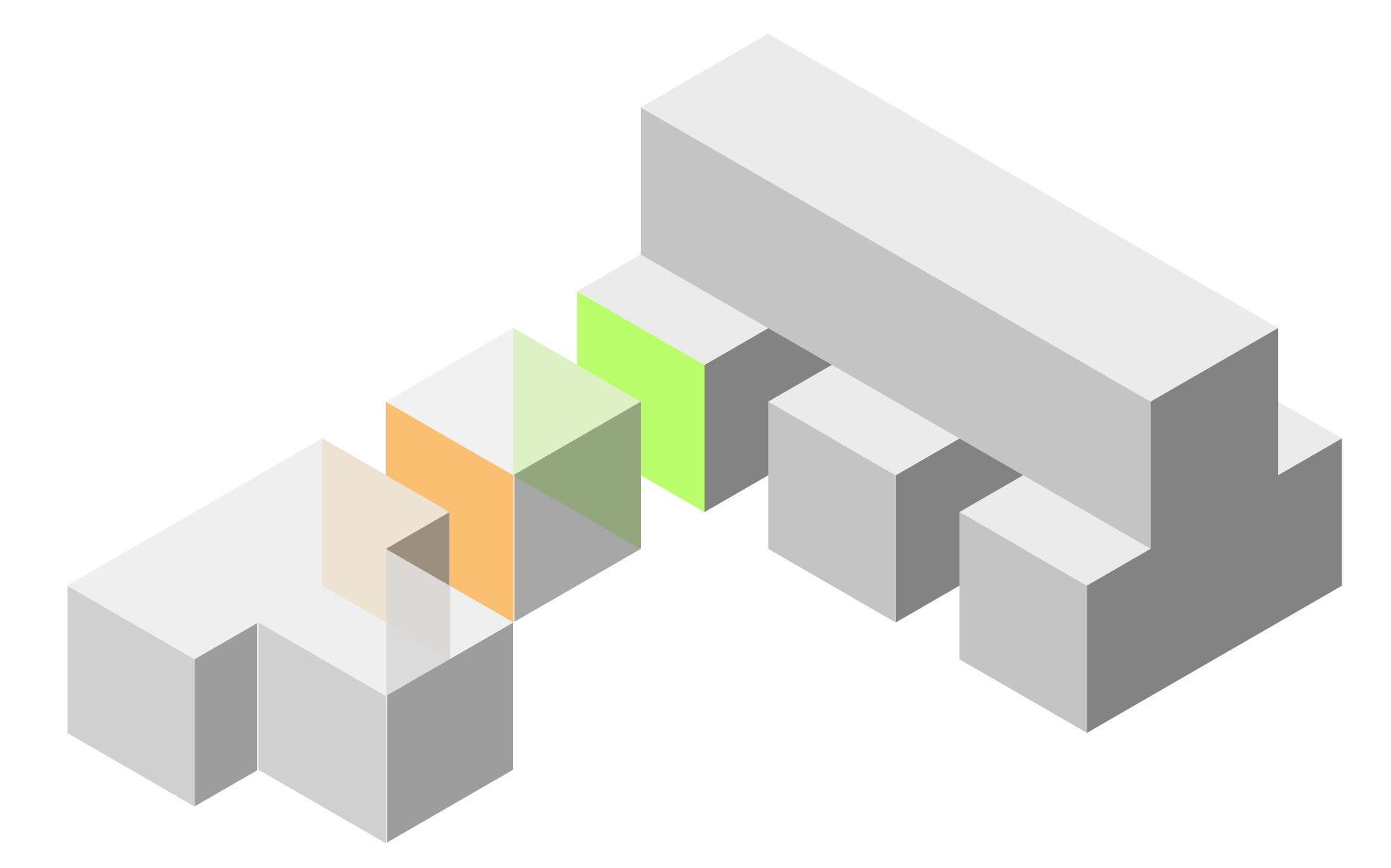}
    \caption{Logical blocks and logical-block network. 
    From left to right, there are the 3-port block, identity, and lattice surgery logical blocks. 
    By matching the ports (colored surfaces) of these blocks, we arrive at a logical-block network.
    }
    \label{fig:logical_blocks}
\end{figure}

\textbf{Surface code fault-tolerant protocols.}
Our primary interest, is on fault-tolerant protocols based on the surface-code. 
In this case, a quantum computation is expressed as a network of logical operations, which we call a \textit{logical block network} following Ref.~\cite{bombin2021logical}. 
Each element in the network is a logical block---an encoded, logical operation realized by a \textit{quantum instrument network} (\textbf{QIN}) along with some input and output ports for the logical information.
Specifics depends on the operations being implemented, as well as the model of computation (CBQC, MBQC, or FBQC)~\cite{flavoursinprep}. 
The native physical instruments used: unitary gates, single qubit measurements, two-qubit parity measurements, resource state generation, and any other operations depend on which model of computation is best suited to the underlying physical hardware.
Blocks are composed along ports to generate larger computations, with a schematic example depicted in Fig.~\ref{fig:logical_blocks}. 
Regardless of the model, the connectivity structure of surface code QINs locally follow a 3-dimensional network representing the space-time history, and both outcomes and topological checks are localized in this structure.
For example, surface code computations using lattice surgery~\cite{horsman2012surface,litinski2019game,bombin2021logical}, or ZX-spider networks~\cite{de2020zx,bombin2021logical,litinski2022active} generate 3D networks like those depicted in Fig.~\ref{fig:logical_blocks}. 
We explain these logical blocks in more detail in Sec.~\ref{sec:scheduling}. 

\section{Modular decoding methods}\label{sec:MDmethods}

A logical block network defines a class of decoding problems $(\checkgen, \membranes, \errorgen)$, with their relation $\partial$ and possibly a probability distribution over $\errorgrp$.
A specific problem instance, is given by the syndrome configuration $\partial \epsilon$ corresponding to a physical error $\epsilon \in \errorgrp$.
In this section, we show how to divide the decoding problem of a network (or part of one) into smaller decoding sub-tasks, which can be solved separately---some in parallel---and then combined to solve the global problem.

\subsection{Decomposing the decoding problem}

Modular decoding acknowledges that the entirety of the decoding problem will only be available once the quantum algorithm is completed.
It addresses the problem of providing logical outcomes throughout the computation  by splitting the monolithic problem into decoding sub-tasks of manageable size.
The solution to each sub-task includes a portion of the global recovery $\kappa$. 
Crucially, these tasks can begin as soon as their input data is available.
Furthermore, as soon as the necessary outcomes and portions of $\kappa$ are available \textit{error corrected logical outcomes} $\bar{v}(M) := v(M)\oplus \kappa_{\membranes}(M)$ can be computed.
This allows the quantum computation to perform feed-forward classical control wherein the logical block structure of the computation changes depending on the extracted classical outcomes.

A {\it recovery-based decoder} proceeds by finding a recovery operator $\kappa \in \errorgrp$, with the property $\partial \kappa = \partial \epsilon$.
The approach taken by modular decoding is to split this problem into decoding sub-tasks indexed by $i \in \dectasks$, with each task committing a portion $\kappa_i$ of the global correction $\kappa$
\begin{align}
\kappa := \sum_{i\in \dectasks} \kappa_i.
\end{align}

How is each component $\kappa_i$ obtained? 
Each decoding task $i \in \dectasks$ is defined by its own set of check generators $\checkgen_i \subseteq \checkgen$, and error generators $\errorgen_i \subseteq \errorgen$.
The check generators $\checkgen_i$ and error generators $\errorgen_i$ defining different decoding tasks can partially overlap in general.

\begin{info}
\begin{itemize}[leftmargin=3ex]
    \item[$\epsilon_i$]$\in \langle \errorgen_i \rangle$: the portion of the physical error configuration   relevant to decoding task $i\in \dectasks$.
    \item[$\mu_i$]$\in \langle \errorgen_i \rangle$: a correction estimate produced by task $i \in \dectasks$.
    \item[$\kappa_i$] $\in \langle C_i \rangle$: the portion of the correction estimate $\mu_i$ committed as part of the global correction $\kappa$ by task $i \in \dectasks$.
\end{itemize}
\end{info}

\begin{figure}[h!]
    \centering
    \includegraphics[width=0.75\linewidth]{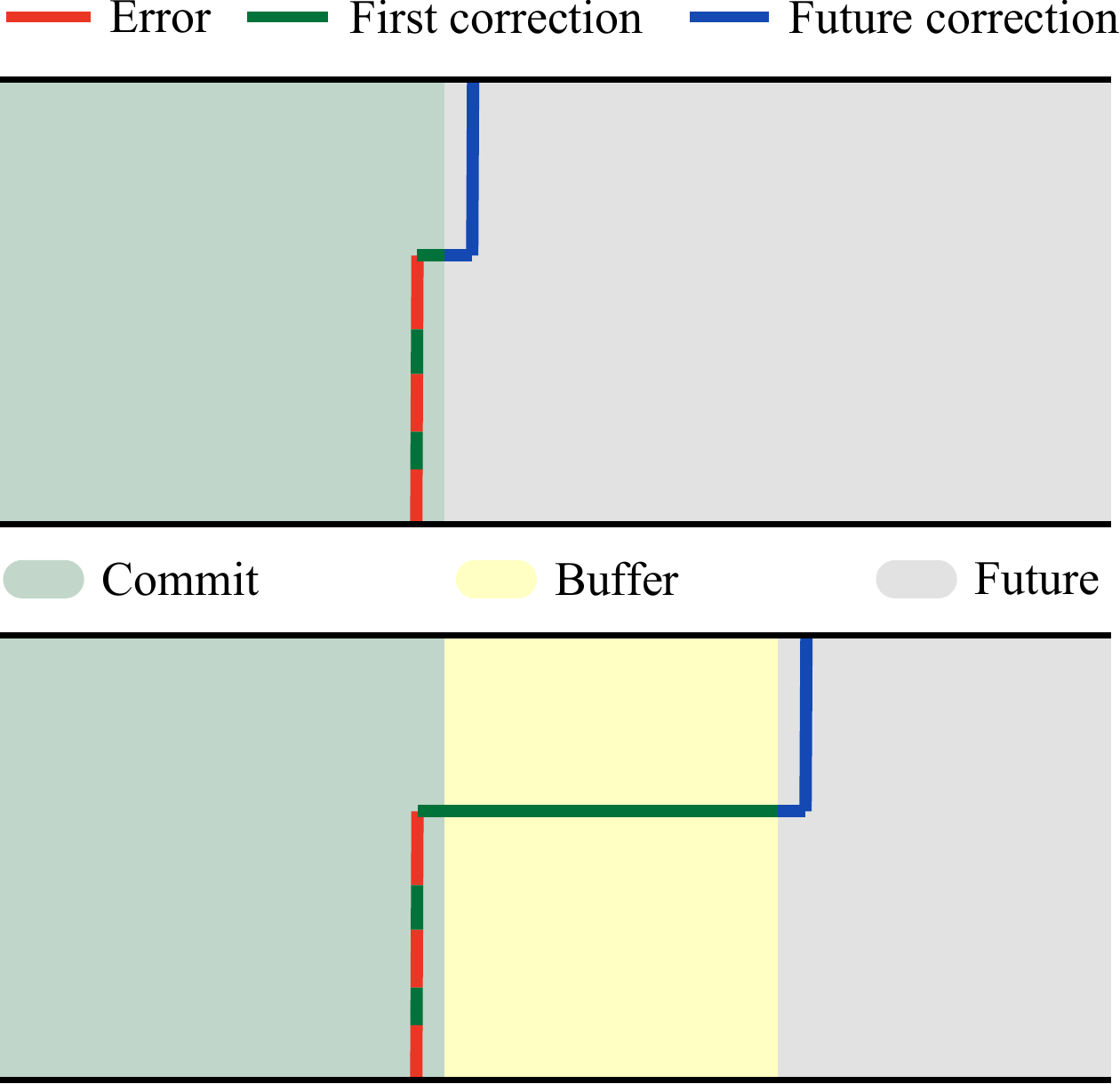}
    \caption{(upper panel) Distance reduction of vanilla modular decoding with no buffer. 
    In a first stage, an error with weight $\approx d/4$ (red) results in an miss-leading partial correction (green line).
    This leaves an updated syndrome which leads further decoding to complete a logical error (blue line). 
    (lower panel) The same error configuration is presented.
    In this case, the first decoder is aware of additional buffer of syndrome information (yellow region).
    A similarly misleading correction (green) is no longer viable for the local decoder as it has  higher weight than the actual error.}
    \label{fig:distance_reduction}
\end{figure}

The result of each decoding task $i$ is a recovery estimate $\mu_i \in \langle \errorgen_i \rangle$.
Crucially, not all of the correction operator $\mu_i$ is taken at face value and committed into the final correction $\kappa$.
Only the restrictions of recovery estimate $\mu_i$ to the smaller commit region $C_i$ are used to determine the final correction $\kappa$.
Having a sufficient buffer is essential to maintaining good decoding performance, as otherwise the low-weight errors may lead to logical faults. 
We refer to this problem as a reduction in the \textit{effective fault distance}. 
Examples of how the effective decoding distance may be halved in the absence of buffers are shown in Fig.~\ref{fig:distance_reduction}. 

The global recovery operator $\kappa$ is obtained by taking portions $\kappa_i$ of these recovery estimates $\mu_i$ (i.e. $\kappa\vert_{C_i} = \kappa_i := {\mu_i}\vert_{C_i}$).\footnote{
    Here, the operator $\vert$  denotes the restriction operator with respect to a specific subset of error generators and assumes a unique representation of the original element being restricted in terms of error generators.
}
The {\it commit regions} $C_i$ partition the full set of error generators $\errorgen$ into disjoint subsets
\begin{align}
    \errorgen = \bigcup_{i\in \dectasks} C_i.
\end{align}
In turn, each commit region $C_i \subseteq \errorgen_i \subseteq \errorgen$ is a subset of the error generators $\errorgen_i$ relevant to decoding task $i$, with
\begin{align}
    \errorgen_i = C_i \cup B_i.
\end{align}
The remaining set of error generators $B_i := \errorgen_i \setminus C_i$ relevant to each decoding task is called the \textit{buffer region} for task $i$ and improves the quality of $\kappa_i$.

\subsection{Data dependency among sub-tasks}
In order to obtain a consistent recovery operation $\kappa$ such that $\partial \kappa = \partial \epsilon$ decoding tasks will need to communicate.
In particular, there are check generators $\sigma \in \checkgen$ which straddle two (or more) commit regions $C_i,  C_j$ (i.e., there exists error generators $e_i \in C_i$ and $e_j \in C_j$ such that $\partial e_i(\sigma) =1 $ and $\partial e_j(\sigma) =1 $). 
Decoding of these tasks will need to coordinate to guarantee $\partial \left( \kappa_i + \kappa_j + \epsilon \right)(\sigma) = 0$.
We call decoding tasks $i, j\in\dectasks$ sharing such a check {\it neighbours} and must somehow communicate.
\footnote{
Note that this notion is dependent on the specific choice of check generators $\checkgen$ and error generators $\errorgen$.
For topological fault-tolerant schemes, there is often a natural set of {\it low-weight} generators which involve a small number of outcomes and detect a small number of error generators.
}

We assume that communication between tasks is exclusively achieved by adapting the input instance of some tasks based on the outputs/solutions obtained for other tasks.
This preserves the functional input-output signature assumed from off-the-shelf decoding algorithms.
This precludes neighbouring decoding tasks from being solved in parallel (otherwise their combined recovery may not satisfy checks straddling both commit regions). 
As such, a sequential causal order is introduced between all neighboring decoding tasks; 
the order of which is a design choice we study in section \ref{sec:scheduling}.

The assumed causal order between decoding tasks can be modelled with a \textit{scheduling graph} $\Gsch := (\dectasks, \prec)$. 
This is a directed, acyclic graph (DAG), with vertex set given by $\dectasks$ (i.e., one vertex per decoding task). 
A directed edge is placed between each pair of neighboring tasks $i \prec j$ with the direction denoting the dependence of the input of $j$ on the output of $i$.
Tasks can only be consistently ordered if the graph is acyclic (i.e., not have directed cycles).
The scheduling graph can partly determine the reaction time---the computational contribution to the reaction time is upper bounded by the depth of the scheduling graph multiplied by the (maximal) time taken for a sub-decoder to return a recovery.
We will discuss scheduling schemes in Sec.~\ref{sec:scheduling}.

\begin{figure}[ht!]
    \centering
    \includegraphics[width=0.85\linewidth]{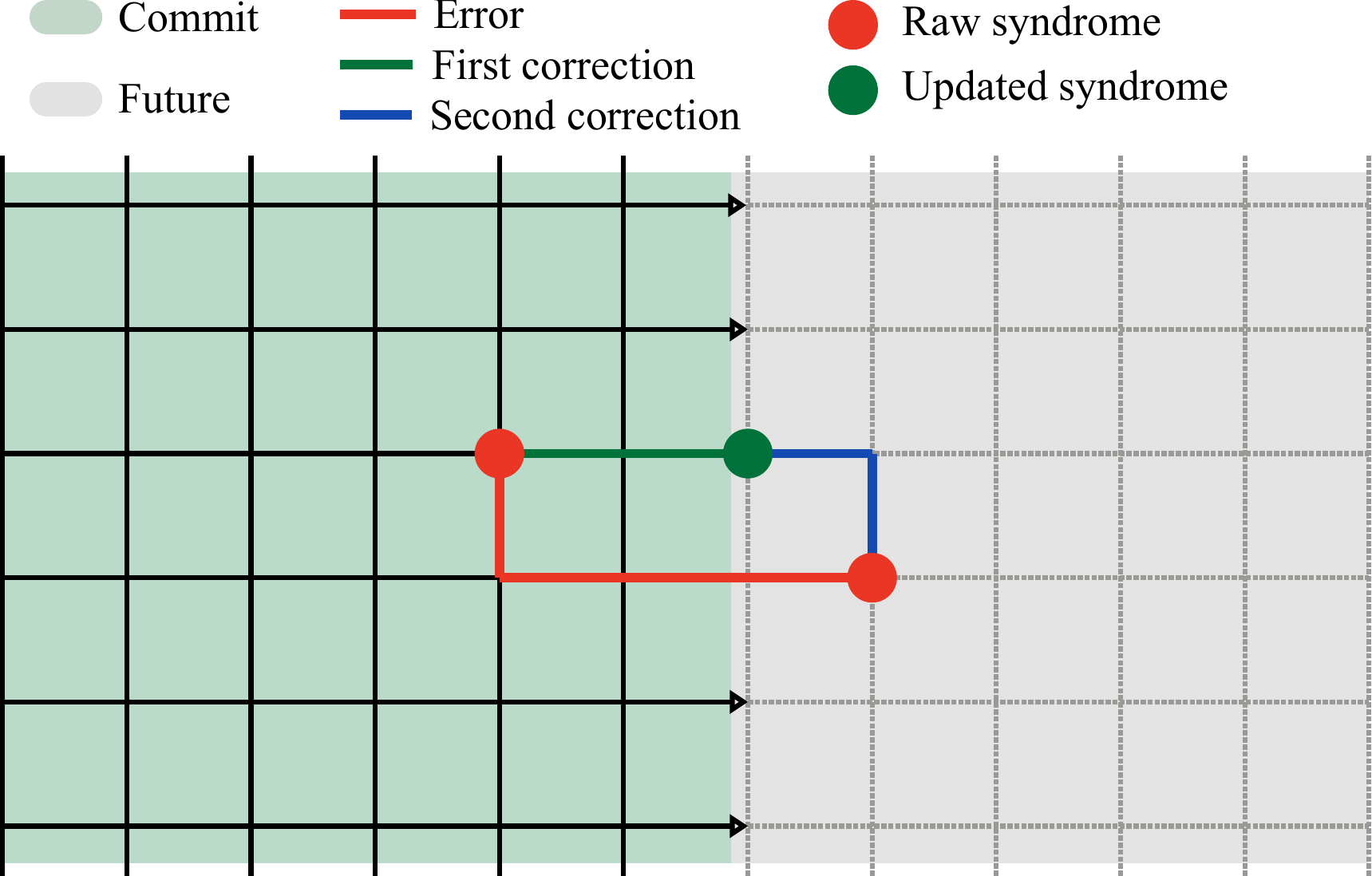}
    \caption{
    The square lattice represents a syndrome graph, with vertices as check generators $\checkgen$ and edges as error generators $\errorgen$ that flip the checks they are incident to.
    The green and grey shading illustrates a partitioning of the decoding problem into two sub-tasks $L$ (left) and $R$ (right).
    Errors $\epsilon$ (red lines) can span multiple components with the syndromes $\partial \epsilon$ (red dots) possibly spread accordingly.
    Error generators at the interface, represented by arrow edges, affect checks in both $\checkgen_L$ and $\checkgen_R$.
    They are included in $C_L$, a subset of the error generators $\errorgen_L$ for the component which is decoded first.
    Doing so guarantees satisfying all check generators $\sigma \in \checkgen_L$ (vertices in the green shaded region) regardless of the future corrections $\kappa_F$.
    However, the committed correction $\kappa_L$ (green line) also produces an \textit{updated syndrome} $\partial (\epsilon + \kappa_L)$ which may be different from $\partial \epsilon$ along the interface, as illustrated by the green dot.
    The second decoding sub-task takes this updated syndrome $\partial (\epsilon + \kappa_L)$ as input to produce a second correction component $\kappa_R$ (represented by the blue line).
    Buffers are not illustrated as they are not necessary to illustrate the notion of residual syndrome and data dependency.
    }
    \label{fig:inter_graph_edges}
\end{figure}

The set of visible syndromes, provided as input for a decoding task $j$, may be altered based on the output from other decoding tasks $i$ preceding it ($i\prec j $).
In particular, instead of the syndrome $\partial \epsilon$ which only includes the effect of the physical errors $\epsilon$, the input instance to decoder $j$ will consist of the syndrome $\partial (\epsilon + \kappa_{P_j})$, where $\kappa_{P_j} := \sum_{i \prec j} \kappa_i$ includes the corrections committed by all prior decoders.
In practice, $\checkgen_j$ will not include check generators which are already guaranteed to be trivial for $\kappa_{P_j} + \epsilon$ and only a small number of check generators in $\checkgen_j$ will be affected by corrections in neighboring tasks and will need to have their syndrome updated in the input to task $j$.
As illustrated in Fig.~\ref{fig:inter_graph_edges}, this can be viewed as setting a boundary condition for task $j$.

\subsection{Logical outcomes from partial membranes and corrections}
For each commit region $C_i$ a sub-decoder commits a fragment $\kappa_i$ of the global recovery $\kappa \equiv \sum_{i \in \dectasks} \kappa_i$.
In particular, the recovery's effect on logical membrane $M$ is given by the map $\kappa_{\membranes}(M) = \sum_{i \in \dectasks}\kappa_{i ~\membranes}(M)$. 
A specific membrane $M \in \membranes$ can only have a finite subset of tasks $\dectasks(M) \subseteq \dectasks$ which can ever contribute to this sum (an intuitive reason for this is that outcomes are not affected by later errors).
We call this subset the relevant tasks to $M$ and denote it by $\dectasks(M) := \{ i \in \dectasks | \exists e\in C_i : e_\membranes(M) \neq 0\}$.
Once all outcomes for $M$ are available and relevant decoding tasks completed, the corrected logical outcome for $M$ can be determined by the (mod~$2$) sum of all partial membranes across the commit regions,  
\begin{align}
\bar{v}(M) := v(M) + \sum_{i \in \dectasks(M)} \kappa_{i \membranes}(M),
\end{align}
where $\bar{v}(M)$ denotes the \textit{error corrected outcome} for $M$.
We will in general assume that the uncorrected outcome $v(M)$ can also be decomposed into partial membrane contributions along a similar partition.

\subsection{Buffer growth \label{sec:buffer_growth}}

To keep the decoding tasks small and reaction time as low as possible, the size of the decoding sub-tasks must be as small as possible without impacting decoding quality.
However, the theorem proved in section \ref{sec:soundness-proof} as well as the numerical result of section \ref{sec:simulations} support the qualitative conclusion drawn from Fig. \ref{fig:distance_reduction} and other examples.
Namely, a buffer region of width $b \geq d$ (or close to it) is needed to maintain the decoding quality of monolithic decoding in a modular decoding approach.
In this section, we describe how minimal buffer regions $B_i := \errorgen_i \setminus C_i$ as well as the set of relevant check generators $\checkgen_i$ for each decoding task $i\in \dectasks$ can be obtained extracted automatically. 
The only input needed is the partition of $\errorgen$ into commit regions $C_i$, the desired buffer distance $b$ and the partial order $\prec$ among decoding tasks.

From each commit region $C_i$, a graph traversal (breadth first search) is performed into other error generators of $\errorgen$.
The graph structure used is the neighbor relation induced by the check generators $\checkgen$.
However, error generators in $P_i$ are not included in the growth phase.
This graph traversal collects all error generators at a graph-distance smaller or equal than a predefined buffer size $b$ and adds them to $B_i$ if they don't already belong to $C_i$.
Since $P_i$ does not participate in the graph traversal, past commits act as a barrier to the buffer growth.
As such, error generators which are a short distance from $C_i$ in the full syndrome graph (or Tanner graph), may end up being excluded due to past commits in $P_i$.

The same growth process can also be used to identify the check generators $\checkgen_i$, which should be actively considered in the decoding task.
The check generators in $\checkgen_i$ are those whose syndrome is fully determined by $P_i \cup C_i \cup B_i$ and not fully determined by $P_i$.
Note that by definition, the growth phase does not proceed into $P_i$ and check generators whose syndromes are fully determined by $P_i$ should already be neutralized by previous commits $\kappa_{P_i}$.

\subsection{Summary: modular decoding anatomy}

We now summarize the various components of a modular decoding problem. 
A global decoding problem can be divided into several, modular decoding sub-tasks, each of which will in general only have access to partial information. 
From the perspective of each sub-task $i \in \dectasks$, the global set of error generators $\errorgen$ is partitioned into four distinct regions.

\begin{info}
\begin{itemize}[leftmargin=3ex]
    \item[$P_i:$] The subset of errors generators ($P_i\subseteq \errorgen$) for which a correction has already been commited in the {\bf past}. 
    \item[$C_i$:] The subset of error generators ($C_i\subseteq \errorgen$) for which the current decoder will {\bf commit} the final correction $\kappa_i$.
    \item[$B_i$:] The current decoder will solve the decoding problem with respect to a larger subset of error generators $\errorgen_i \equiv C_i \cup B_i$.
    The subset $B_i$ is treated as a {\bf buffer} to improve the decoding quality in $C_i$.
    The $B_i$ component of the correction estimate $\mu_i$ will be discarded or revised later.
    \item[$F_i$:] The subset of error generators ($F_i\subseteq \errorgen$) from the {\bf future},  which have no influence on visible check generators.
\end{itemize}
\end{info}

The decomposition into these components is central in the soundness proof of modular decoding provided in section \ref{sec:soundness-proof}.
Our approach to specifying the decomposition of modular decoding into sub-tasks will be to start from a partition of the error generators $\errorgen$ into commit subsets $C_i$.
To do so, we will mirror the logical block decomposition of the circuit.
At this point one of a few sensible schedules $\Gsch$ discussed in section \ref{sec:scheduling} can be chosen to provide a data dependency structure among decoding tasks.
Finally, the $P_i$, and the buffer regions $B_i$ can be obtained algorithmically given a target buffer size $b$ (see \ref{sec:buffer_growth}),  which should be chosen as $b \approx d$.

In addition to this partition of the error generators, it is sometimes necessary to connect to the actual set of physical measurement outcomes $V_i$, which are in principle available to a given decoding task.
\begin{info}
\begin{itemize}[leftmargin=3ex]
    \item[$V_i$:] The {\bf visible} subset of outcomes ($V_i\subseteq \outcomes$) which are {\it in principle} available to the current decoder task $i$.
    These typically includes past outcomes which are no longer directly relevant to the current decoding task.
\end{itemize}
\end{info}

In practice, the input to a decoding task instance is based on syndromes and syndromes are based on outcomes.
It will be sufficient to assume that $V_i$ is the smallest subset of outcomes $\outcomes$ supporting the check generators in $\checkgen_i$ and any $\checkgen_j$ with $j\prec i$.
In practice however, many of the check generators in $\checkgen_{V_i}$ are already guaranteed by $\kappa_{P_i}$ and are not influenced by error generators in $B_i \cup C_i$ so it is sufficient for the local decoding task to focus on $\checkgen_i$. 

When we are dealing with a setting where there is a one to one correspondence between error generators and outcomes, the {\it visible} region is specified by $V_i = P_i \cup C_i \cup B_i$.
Based on our algorithmic construction of $B_i$ and $\checkgen_i$ from Sec.~\ref{sec:buffer_growth}, we may interpret $\checkgen_{V_i}$ to be a maximal subset of check generators unaffected by error generators in $F_i$, and $V_i$ the minimal subset of outcomes supporting said checks.
The practical importance of $V_i$, is that the decoder unit responsible for task $i$ needs to wait for said outcomes to be available before it can begin solving its task.

\section{Modular decoding with buffers: proof of decoder soundness \label{sec:soundness-proof}}

In this section, we prove that modular decoding is sound.
Namely, we show that, under two assumptions, it can achieve an effective fault distance $d$ equal to the fault distance of the original decoding problem.

The first assumption, is local {\bf decoder soundness}, which is a requirement on the base decoding algorithm used to obtain corrections for individual decoding sub-tasks, as was defined in Def.~\ref{def:decsound}.
We recall that some decoders in the literature such as MWPM~\cite{dennis2002topological,kolmogorov2009blossom} and UF~\cite{ delfosse2017almost} satisfy this property; these are good candidates to use as local decoders.

\begin{figure}[!ht]
    \centering
    \includegraphics[width=0.95\linewidth]{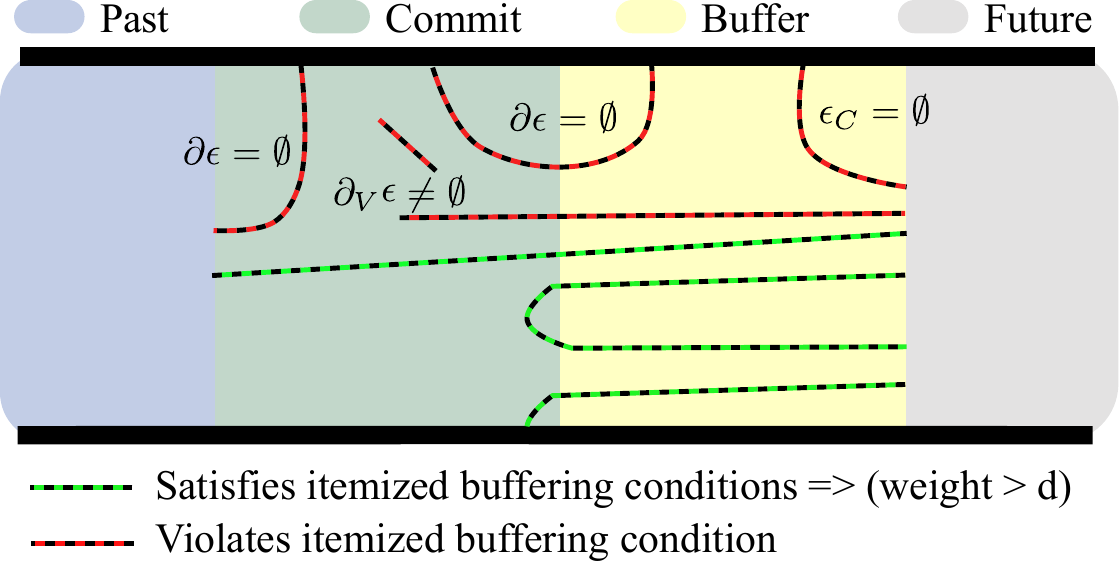}
    \caption{
    The figure illustrates a collection of error strings (connected error clusters) supported on $B\cup C$.
    The buffering condition (Def.~\ref{def:buffering_condition}) states that low weight error clusters satisfying certain {\it itemized requirements} should not exists. 
    Error clusters which satisfy itemized conditions (dashed green black), must have weight larger than $d$ for the buffering condition to be satisfied.    
    The ones that do not satisfy itemized conditions (dashed red black) may have lower weight.
    For these, formulas identify which of the three {\it itemized requirements} is not fulfilled. 
    }
    \label{fig:BufferingCondition}
\end{figure}

The second assumption, constrains the size/shape of the commit and buffer regions used to define individual decoding sub-tasks.
It goes by the name of {\it buffering condition}, and is stated as follows:

\begin{defn}\label{def:buffering_condition}[{\bf Buffering condition}]
For a decoding sub-task with buffer, any connected errors $\epsilon$ satisfying the following conditions will have weight $d$ or larger.
\begin{itemize}
    \item $\epsilon$ is supported on $C \cup B$
    \item $\epsilon$ is undetectable in the visible region ($\partial_V \epsilon = \emptyset$).
    \item $\epsilon$ has support in the commit region ($\epsilon_C \neq \emptyset$).
    \item $\epsilon$ is detectable globally ($\partial \epsilon \neq \emptyset$).
\end{itemize}
\end{defn}
See Fig.~\ref{fig:BufferingCondition} for examples of error clusters which satisfy (or not) the itemized buffering conditions. 

Note that throughout this section, the index $i \in \dectasks$ for the current decoding task is kept implicit throughout the inductive proof.
We nevertheless use the labels $P$, $C$, $B$ and $F$  as in section \ref{sec:MDmethods}, to refer to the partition of error generators $\errorgen$ into past, commit, buffer and future from the perspective of $i$.
The decoder in charge of a decoding task has access to outcomes in $V \subseteq \outcomes$ at best.
Consequently, instead of the full syndrome \(\partial\) it only has access to a partial syndrome
 \(\partial_V\) corresponding to those check generators supported exclusively on visible outcomes $V$.
We will assume that $V$ is the minimal set of outcomes supporting all check generators which are unaffected by errors in $F$.
In other words, $V$ supports all check generators $\sigma \in \checkgen$ whose syndrome is fully determined by the error restriction to $P \cup C \cup B$.

While the buffering condition is expressed in terms of an arbitrary error $\epsilon$, in the soundness proof the condition is applied to specific combinations of errors and corrections.
These combinations are constructed to be ``locally neutral'' as they combine a decoder-generated recovery operation with its instigating error.

The buffer growth method in Sec.~\ref{sec:buffer_growth} (with $b :=d$) generates decoding regions which satisfy the buffering condition.
If we use $d$ as the buffer depth, then no error cluster with weight smaller than $d$ can have support on both the commit $C$ and future $F$, because errors from these two regions have graph-distance larger than $d$.
This is guaranteed by the graph traversal approach use to determine $B$.\footnote{
    The graph structure used is derived from the elementary error generators of $\errorgen \setminus P$ and the elementary check generators $\checkgen$.
}

\begin{thm}[Modular decoding theorem]
If at every step of modular decoding (i.e., for every decoding sub-task) the buffering condition and local decoder soundness are satisfied, then the overall modular decoding procedure satisfies the soundness condition.
\end{thm}

{\bf Informal argument}: Intuitively, the buffering condition means that any {\it connected} error cluster with weight smaller than $d/2$ which is partially supported on the commit region, ($C$) must be fully supported on commit + buffer ($C\cup B$). 
Because otherwise a ``round trip'' of this string that is undetectable in commit + buffer, has two open ends in the (non-buffered) future, and has weight smaller than $d$ -- violating the buffering condition. 
As a result, the local decoder will provide the correct recovery (say, a string from points $x$ to $z$), but only apply the component in the commit region (a string from points $x$ to $y$). The missing part (a string from $y$ to $z$) will be completed by the following sub-decoders, because if the minimum weight path from $x$ to $z$ is through $y$, then the minimum weight path from $y$ to $z$ has the same weight as the $yz$-segment on the $xy$-path. \\

We will continue to use Greek letters to describe error strings and correction candidates.
These will further proliferate in the proof of modular decoding soundness so we preemptively summarize their interpretations in table \ref{key:greek-symbols}.

\begin{table}[htp!]
\caption{\textbf{Symbol key: errors and recovery operators}}
\label{key:greek-symbols} 
\begin{info}
\begin{itemize}[leftmargin=3ex]
    \item[$\epsilon$:] the physical error configuration.
    \item[$\kappa$:] the global correction the modular decoder commits to.
    \item[$\mu$:] the correction estimate proposed for the current decoding task $C\cup B$. ($\mu \equiv \mu_{CB}$). ($\kappa_C \equiv \mu_C$).
    \item[$\tau$:] a processed portion of the errors $\epsilon$ reliably addressed by the previous correction estimates. ($\tau \subseteq \epsilon$).  ($\tau_P \equiv \epsilon_P$).
    \item[$\nu$:] a viable low-weight correction candidate for $\tau$ compatible with previous commits (ignores $\bar{\tau}$).
    \item[$\bar{\tau}$:] an unprocessed portion $\bar{\tau} := \epsilon - \tau$ of the errors $\epsilon$. 
\end{itemize}
\end{info}
\end{table}

\begin{figure}[ht!]
    \centering
    \includegraphics[width=0.95\linewidth]{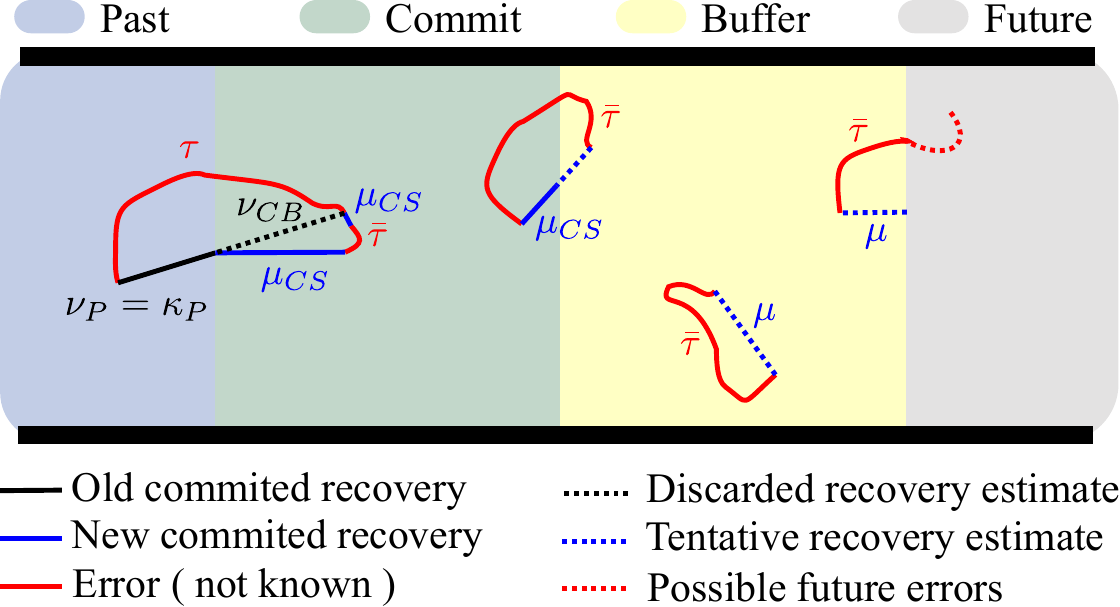}
    \caption{
    An illustrative configuration of the different regions as well as error and correction chains.
    Errors and corrections are labeled by Greek letters following Sec.~\ref{key:greek-symbols}
    Here, the {\it decoding causal order} is presented from left to right whereas in general the order is decoupled from the particular geometric coordinates of the decoding problem.
    The error $\epsilon$ (red lines) is partitioned into a processed component $\tau$, which can be compensated by a lower weight correction $\nu=\nu_{P}+\nu_{CB}$, and the unprocessed error $\bar{\tau} = \epsilon - \tau$. 
    The unprocessed error $\bar{\tau}$ affecting visible outcomes can be decomposed into connected components, some of which will share support with the current commit region $C$. 
    The recovery estimate $\mu$ produced for the current decoding task contains $\mu_{CS}$, the parts which the proof deems as necessarily trustworthy, including the correction $\mu_C :=\mu_{CS}\cap C$ to be committed.
    Sub-optimal correction estimates in parts of $\mu$ disconnected from $C$ are less harmful to the global decoding quality, as these will be revised with the benefit of future information.
    Note after this step, the past region $P$ grows to include the current commit $C$, and the committed correction $\kappa_P$ will be extended with $\mu_C$. 
    }
    \label{fig:proof-guide}
\end{figure}

Whereas the proof applies generally, to stabilizer fault-tolerance with some notion of locality, our intuition is derived from syndrome graph type decoding problems, 
for which Fig. \ref{fig:proof-guide} can further illustrate the elements involved in the proof.

In the context of the current decoding task, $\kappa_P$ is already defined and corresponds to past committed error estimates.
The available syndromes of the error are given by $\partial_V \epsilon$. 
The decoder returns a correction estimate \(\mu\) with
\[
\mu_{CB}\equiv \mu,\qquad \partial_V\mu= \partial_V(\kappa_P + \epsilon_V). 
\]
The global correction is committed for region  $C$ as \(\kappa_C := \mu_C\).

Notice that the decoder might need to \textbf{abort} if there is no error estimate \(\mu\) which is local to $C\cup B$, \(\mu \equiv \mu_{CB}\), and is compatible with the error $\epsilon$ and previously committed estimate $\kappa_P$,
\(\partial_V \mu = \partial_V (\kappa_P + \epsilon) \). 
This is also a decoding failure, and is taken into account below.\footnote{
    In practice, the syndrome for $\checkgen_i \subseteq \checkgen_V$, which excludes checks generators already guaranteed by $\kappa_P$ is enough (see Sec. \ref{sec:MDmethods}).
    Since $i \in \dectasks$ is kept implicit here, we use $\checkgen_V$ to distinguish from $\checkgen$.
}

\textbf{Proof.} 
The strategy is to show inductively that, assuming \(|\epsilon|<d/2\),
at every step there exist \(\tau\) and \(\nu\) such that the following condition is satisfied\footnote{Notation: binary vectors are identified with set through indicator vectors.}
,
\begin{align*}
\tau\subseteq \epsilon,\quad \tau_P =\epsilon_P, \quad \nu_P=\kappa_P, \quad
|\nu|\leq |\tau|,\quad \partial \nu=\partial \tau.
\end{align*}

Within the inductive proof, $\tau$ can be considered as a fragment of the error $\epsilon$ already dealt with.
Conversely, $\nu$ is a viable extension of the committed correction $\kappa_P$ into a viable global correction for  $\epsilon$.
Neither of these have a physical (or programmatic) counterpart, and only play a role in the proof.

This is enough because when it holds after the last step then
\(|\kappa|\leq|\epsilon|\).
Each inductive step corresponds to a decoding task which is performed, and the order in which such steps are taken in the proof can be any complete order compatible with the causality relation $\prec$ imposed on decoding tasks.

For the base case of the induction, it suffices to take \(\tau=\nu=0\). For the inductive
step, assume that indeed we have such \(\nu\), \(\tau\). 
The aim is to construct some \(\nu'\), \(\tau'\) satisfying the required conditions
after the modular decoding step (i.e., $P \mapsto P \cup C$). 
Let
\[
\bar \tau :=\epsilon-\tau.
\]
The modular decoding step produces some \(\mu=\mu_{CB}\) satisfying
\begin{align*}
\partial_V \mu =& \partial_V (\kappa_P +\epsilon_V)
= \partial_V (\nu_P + \tau_P + \epsilon_{CB} ) \\
=& \partial_V (\nu_{CB} +\tau_{CB} + \epsilon_{CB})
= \partial_V (\nu_{CB} + \bar\tau_{CB}).
\end{align*}
(The existence of \(\mu\) is a consequence of the rightmost expression.)
The correction \(\mu\) has minimal weight by virtue of {\it decoder soundness}, since
\[
|\bar\tau_{CB} + \nu_{CB}|\leq |\bar \tau|+|\nu|\leq |\bar \tau|+|\tau|=|\epsilon|\leq d/2.
\]
Consider a region \(S\) obtained as the union of the {\it support} of those connected
components of
\[
\alpha:=\mu + \bar\tau_{CB} + \nu_{CB}
\]
that contain some element of the commit region $C$. 
Since \(\partial_V \alpha = 0\) we have \(\partial_V \alpha_S = 0\), that is
\[
\partial_V \mu_S = \partial_V (\bar\tau_S + \nu_S).
\]
and \(\mu_S\) has to be optimal itself, so that
\[
|\mu_S|\leq |\bar \tau_S|+|\nu_S|,
\]
The buffering condition applies to each connected component of
\(\alpha_S\) and thus \(\partial\alpha_S=0\), i.e.,
\[
\partial \mu_S = \partial(\bar\tau_S + \nu_S).
\]
Let \(C\cup S\) be the region obtained as the union of \(S\) and region $C$.
Clearly \(\alpha_S = \alpha_{CS}\) and thus
\[
|\mu_{CS}|\leq |\bar \tau_{CS}|+|\nu_{CS}|,
\qquad\partial \mu_{CS} = \partial(\bar\tau_{CS}+\nu_{CS}).
\]
We take
\[
\nu':= \nu+\nu_{CS}+\mu_{CS},\qquad \tau':=\tau+\bar \tau_{CS}
\]
and it suffices to check that
\begin{align*}
\tau'\subseteq \epsilon, \quad
\tau'_{C} = \epsilon_{C}, \quad 
\nu'_C = \kappa_C,\\
|\nu'|\leq |\tau'|,\quad 
\partial \nu' = \partial \tau'.
\end{align*}
Indeed:
\begin{align*}
\tau' &=\tau+\bar\tau_{CS}\subseteq \tau\cup\bar\tau=\epsilon,\\
\tau'_{C} &= \tau_{C}+(\bar\tau_{CS})_C = \tau_C+\bar\tau_C =\epsilon_C, \\
 \nu'_C &=\nu_C+(\nu_{CS})_C+(\mu_{CS})_C\\
      &=\nu_C+\nu_C+\mu_C=\mu_C=\kappa_C,\\
|\nu'| &\leq|\nu|-|\nu_{CS}|+|\mu_{CS}|\leq |\tau|+|\bar \tau_{CS}|= |\tau'|,\\
\partial \nu' &= \partial(\nu+\nu_{CS} + \mu_{CS}) = \partial\tau+\partial\bar \tau_{CS}= \partial \tau'. \qed
\end{align*}

\hypertarget{erasure}{%
\subsection{Modular decoding with erasures}\label{sec:erasure}}

In a Pauli noise model each error in $\errorgrp$ occurs according to a certain probability distribution (usually described by an \textit{i.i.d.} model over the generators $\errorgen$).
The control software must get any information about which error actually happened from the extracted syndrome.
In contrast, for an error model including erasures, there is additional information available.
Rather than having an outcome be flipped with some probability and only be able to infer whether the outcome was flipped or not from syndrome information, some outcome may be  flagged as \textit{erased}.
This information is provided by \textit{herald outcomes}, which go beyond the $Z_2$ linear structure of checks and logical outcomes.
Rather than a probability distributions over Pauli faults, 
a noise model including erasure can be seen as a conditional distributions of Pauli faults conditioned on the sample drawn from a probability distribution of erasures.

Traditionally, each erasure correspond to a missing measurement outcome.
An outcome $o$ which is flagged as erased may be assigned an arbitrary \textit{guess value} $v(o)$.
This corresponds to having an error generator $e_o \in \errorgen$, which corresponds to a measurement error on $o$ and is modeled to occur with $50\%$ probability whenever the erasure is heralded for $o$.

A decoder which is aware of an erasure flag on $o$, will not take the value $v(o)$ seriously.
In fact, this value may be missing from the input and can be generated at random by the decoder itself, and reassign its value if the initial guess is otherwise deemed incompatible with the most likely error configuration.

In this sense, erasure is a more benign form of noise than flip noise since an outcome flip probability $p/2$ is equivalent to an outcome erasure probability of $p$ where the herald information (i.e., which outcomes were erased) is ignored.
Furthermore, if erasures are the only form of noise, exact decoding can be performed wherein logical outcomes are either correctly recovered or lost in a heralded way.
A fault tolerant stabilizer protocol which has fault distance $d$ w.r.t. a Pauli error model will also be able to perfectly recover from up to $d-1$ erasures w.r.t. the same error generators.

We may apply the soundness proof of modular decoding to a \textit{post-erasure} fault-tolerance protocol. 
In other words, we can seek to partition a decoding problem on which erasures have already been taken into account.
Once the erasures are taken into account one is left with a decoding problem with distance $d'\leq  d$ the conditional Pauli error distribution over a subset of the remaining error generators.
The modular decoding theorem need only be applied to distance $d'$, requiring less buffering.
However, the decoding problem resulting from including the erasure instance will also be \textit{less local}. Check generators affected by a shared erasure will be considered as being at distance zero from each other.

Applying the modular decoding theorem to a decoding problem where the erasures have already been fixed (sampled) suggest that the partitioning into sub-tasks may also be taken to depend on the observed erasures.
Constructing the buffer regions in a way which depends on the specific erasure configuration drawn makes intuitive sense.
Buffer regions in parts of the protocol with an atypically high proportion of erasures will need to be bigger due to the faster buffer growth through erasure clusters.
Conversely, buffer regions with a low proportion of erasure can be kept smaller, possibly reducing to a buffer width of $d'$ for a fragment with no erasures.
This suggest a heuristic by which to minimize the size of the needed buffers (or increase the effectiveness of fixed size ones).
One should attempt to partition the decoding problem along cuts such that the immediate buffer regions have the smallest density of erasures.

\section{Scheduling decoding tasks}\label{sec:scheduling}

In this section we define several approaches to modular decoding. 
In particular, we define a schedule of commit regions and their associated buffers, that determine the set of sub-tasks and their communication requirements. 
We include a schematic for the system-level data flow requirements of these implementations in App.~\ref{app:system_data_flow}.

\textbf{Logical blocks, ports, and membranes.}
To specify our schemes, we first need to carefully define the constituents of a surface-code logical space-time network, known as \textit{logical blocks}~\cite{bombin2021logical}. 
A logical block is a set of instructions to implement a logical (i.e., encoded) quantum instrument based on surface codes. 
These specify the physical instructions that must be implemented to manipulate topological defects of the code, such as primal and dual boundaries~\cite{kitaev2006anyons,preskill1999lecture, raussendorf2007topological, raussendorf2007fault}, transparent domain walls, twists and corners~\cite{bombin2009quantum, bombin2010topological, barkeshli2013twist, levin2013protected, liu2017quon} in order to realize the desired logical operation. 
In addition to these topological features, logical-blocks have a set of \textit{ports}, which correspond to the inputs or outputs of the (logical) quantum instrument and carry information encoded using surface-codes. 
Logical blocks can be composed along ports to build larger logical networks describing fault-tolerant quantum computations.

The connectivity of a logical block network can be described in a simplified way by a directed-acyclic graph (DAG) $\Glog$ in which (i) vertices are logical blocks describing a logical quantum instrument and (ii) edges are ports representing a logical quantum system (here, a single logical qubit in a surface code). 

The directed edges of $\Glog$ define the order in which the logical blocks are performed.
In particular, one can label the vertices with integers such that edges point from the vertex with the smallest label to that with the largest. 
Applying the (logical) quantum instruments sequentially according to this ordering gives a mapping from a collection of logical subsystems to a collection of logical subsystems. 
As we focus on networks comprised of Clifford quantum instruments, we can describe the network with the stabilizer formalism~\cite{bombin2021logical,flavoursinprep}. 
In particular, one can define a set of Pauli operators that stabilize each (logical) quantum instrument. 
Each vertex of the logical block network produces a set of classical outcomes that determines the signs the Pauli operator that stabilize the instrument (i.e., a partial Pauli frame). 
The specific set of physical outcomes that determine the logical Pauli frame are supported on a logical membranes. 

\begin{figure*}[ht!]
\includegraphics[width=0.99\linewidth]{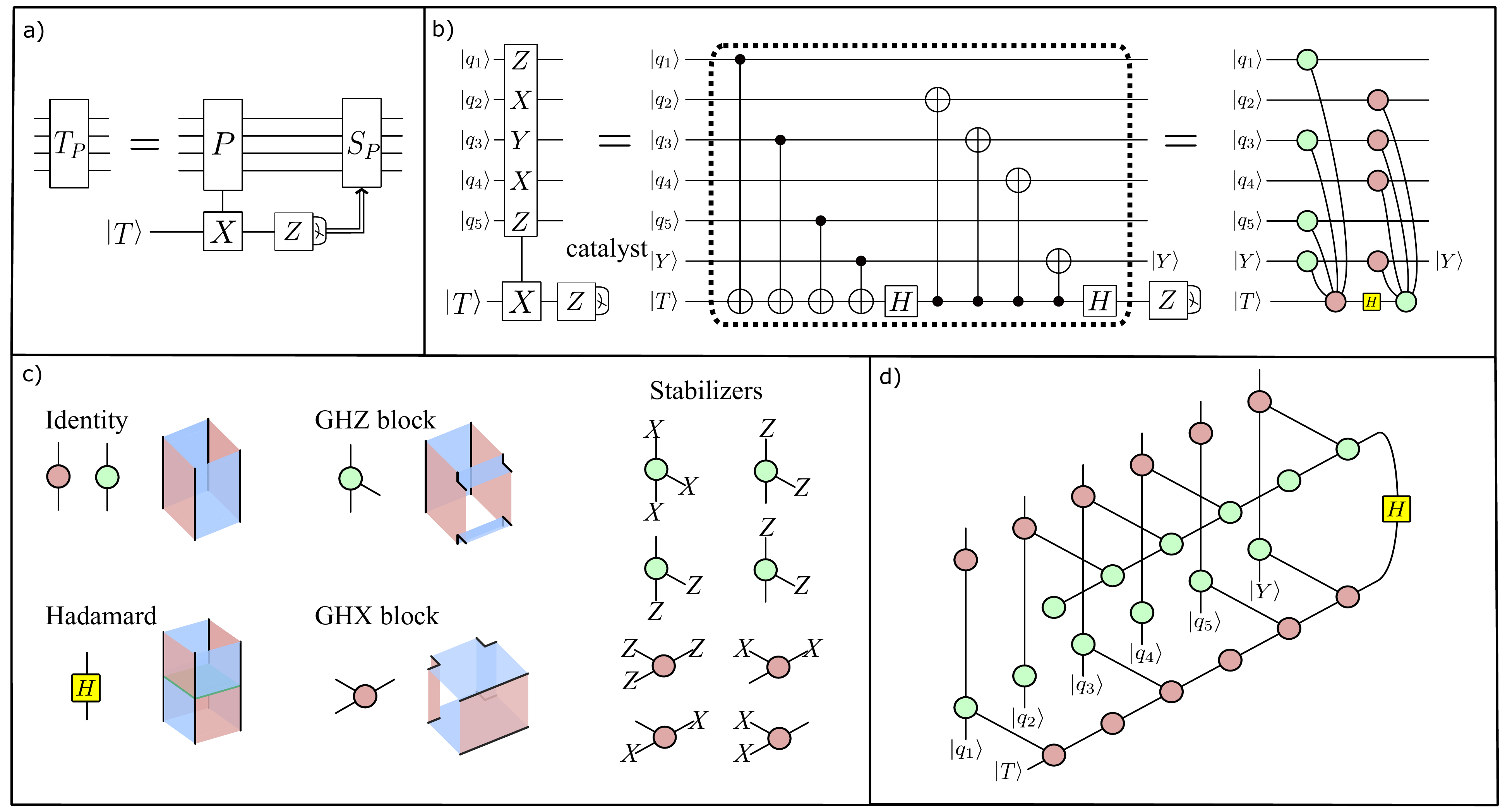}
\caption{
\textbf{(a)} 
A circuit identity showing how an arbitrary $n$-qubit $T_P := \exp{\left(i \pi/8 P\right)}$ gate, for an $n$-qubit Pauli product operator $P$, can be realized by the inclusion of a magic state, along with a generalized CNOT operator controlled on $P$ and conditional $S_P := \exp{\left(i \pi/4 P\right)}$ gate conditioned on a single qubit measurement on the injection register. 
The conditional $S_P$ can be accounted for by the Clifford frame tracking in the fault tolerant setting. 
\textbf{(b)} 
The generalized CNOT conditioned on $P$  can be realized (dashed box) by a circuit involving sequence of CNOTs, and two Hadamard operators.
The decomposition uses a $\ket{Y}$ state (i.e. a $Y$ eigenstate) which is kept unchanged in the process (a catalyst state).
The circuit structure can be simplified into a ZX diagram.
The figure gives the example of a 5-qubit Pauli operator $P = ZXYXZ$.
\textbf{(c)} Each element of the graph $\Glog$ is realized by a fault-tolerant logical block. 
Stabilizers describing the quantum instrument are shown, each one corresponding to a membrane in the logical block. 
The blue and red boundaries correspond to primal and dual boundaries, following the conventions of Ref.~\cite{bombin2021logical}. 
\textbf{(d)} A ZX-diagram equivalent to the one obtained for the 5-qubit example in (b) compatible with elementary logical blocks arranged in a local 3D structure.
This is the layout prescribed by the CNOT matrix architecture.
}
\label{fig:LogicBlocks}
\end{figure*}

Examples of logical blocks are shown in Fig.~\ref{fig:LogicBlocks}. We will consider a class of networks where the (logical) quantum instruments are constructed from ZX-networks~\cite{coecke2008interacting,van2020zx,de2020zx} which we call ZX-instruments following Ref.~\cite{flavoursinprep}. Logical blocks realizing these ZX-instruments are shown in Fig.~\ref{fig:LogicBlocks}. 
We note that as the instruments are Clifford, we may also change the direction of any edges and have a valid logical block network -- as such we often do not draw the arrows explicitly. 
For a given logical block (or network), we label ports by integers, and use them to index the logical correlators of the network. 

\textbf{The CNOT matrix architecture.}
To demonstrate the universality of the decoding schemes, we present a  universal architecture based on ZX-instruments. 
The scheme is based on that proposed in Ref.~\cite{kim2021faulttolerant}. 
The architecture can also be understood in the lattice surgery perspective~\cite{litinski2019game}, however, this presentation will lead to a logic block network that directly admits a fast modular decoding scheme (as demonstrated in the following subsection). 

A universal set of gates on $n$ qubits is given by $\{T_P=\exp \left(\frac{i \pi}{8} P \right) ~|~ P \in \mathcal{P}_n\}$, where $\mathcal{P}_n$ is the $n$-qubit Pauli group. 
Similar to a $T$ gate, each gate $T_P$ can be realized by injecting a magic state and applying a generalized CNOT operator before measuring out the auxiliary qubit.
The generalized CNOT operator with source Pauli product operator $P$ and target qubit having a Pauli operator $X$ is given by $\text{C}_P\text{NOT} = ( 1 + P + X - PX)/2$. 
A conditional Clifford correction $S_P=\exp \left(\frac{i \pi}{4} P \right)$ must be applied depending on the measurement outcome, as shown in Fig.~\ref{fig:LogicBlocks} a). 
In other words, the generalized CNOT operator, measurements and a supply of magic states is a universal set of operations. 
Using an additional $Y$-eigenstate ancilla as a catalyst (that is not consumed) it is possible to implement the generalized CNOT for an arbitrary Pauli operator $P$ using only elementary CNOT operations and two Hadamard operations, as shown in Fig.~\ref{fig:LogicBlocks} b).
This circuit can be represented as a logic block network with each element being a ZX-instrument as shown in Fig.~\ref{fig:LogicBlocks} which we refer to as the CNOT matrix architecture. 
Each element of the network is a ZX-instrument with a space-time volume of $d^3$. 
See Refs.~\cite{bombin2021logical,flavoursinprep} for more details. 
This architecture will be used to illustrate the modular decoding schemes.

\textbf{From logical block networks to scheduling graphs.}
In the following, we use the graph $\Glog$ describing the logical block network to determine the scheduling graph $\Gsch$, governing the set of sub-decoding problems and their order. 
We make this choice for convenience and note that one may make other choices of scheduling graphs. 
For example, nodes of the scheduling graph may be decomposed or coarse grained (just as logical blocks can be) to define alternate decompositions of the decoding problem.
It is desirable for the elementary logical blocks at each vertex to have small volume, such that the processing time for each decoding task is short. 
The decomposition of fault-tolerant logic into networks of logic block satisfy this desiderata, with each block having a volume of $O(d^3)$.

\subsection{Vertex-only decoding}
\begin{figure}[!ht]
    \centering
    \includegraphics[width=0.8\linewidth]{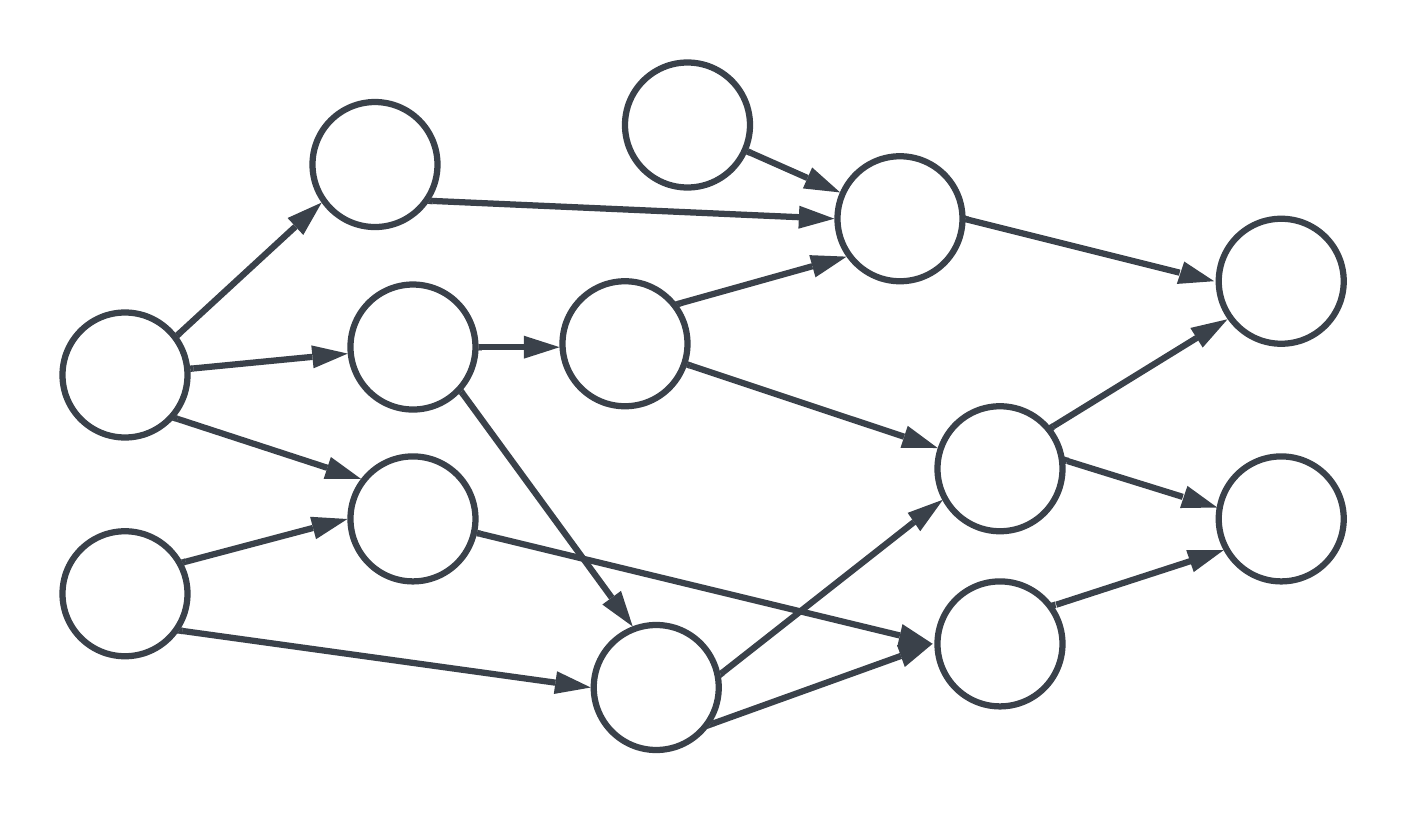}
    \caption{Directed acyclic graph (DAG) $\Gsch$. 
    In our context, each node represents a decoding task $i \in \dectasks$, and each directed edge represents a causal dependence $i \prec j$ imposed whenever the correction $\kappa_i$ committed by an earlier task affects the syndrome for check generators $\checkgen_j$ used by a later task.
    The direction of the edge points from the earlier decoding task to the later one w.r.t. the scheduling.}
    \label{fig:dag}
\end{figure}

The first set of schedules we consider, directly obtain the vertices and edges of the scheduling graph from the logical block network. 
In particular, every logical block (corresponding to a vertex in $\Glog$) defines the commit region (corresponding to a vertex in $\Gsch$) of a sub-decoding problem, and edges are placed between two blocks whenever they share a port. 
What remains is to determine the directions $\prec$ of the edges, the order of the decoding sub-tasks which ultimately impacts the reaction time.
This is schematically represented in Fig.~\ref{fig:dag}. 
The buffers for each region can be chosen by considering the neighbouring blocks. 
One may take all neighbouring blocks connected to the a given block as the buffer, or alternatively use the graph traversal approach described in Sec.~\ref{sec:buffer_growth} to construct the minimum necessary buffers. 

\textbf{Sequential vertex decoding.}
The simplest schedule is obtained by completely mirroring the logical order in $\Glog$ in the dependencies and scheduling of decoding tasks making $\Gsch$ identical to $\Glog$. 
This is called sequential decoding---the sub-tasks proceed in the same order as the logical blocks are performed. 
If decoder modules can process each decoding block faster than the time it takes a block of input outcome data to be generated by the quantum hardware, this naive approach works fine.
In fact, it minimizes reaction time as each block can set the boundary condition for its future neighbours.

However, this approach will generate backlog even if the individual decoder modules take slightly longer to solve decoding tasks than it takes the quantum hardware to produce a block of outcome data.
This is true, even if additional decoding power is made available in the form of additional decoder modules.
This is a poor paralelization of the global decoding problem and decoder modules will lay idle waiting for other modules to complete their tasks which are needed for input boundary conditions.
The data dependency among decoding tasks leads to an accumulating backlog and increasing reaction time with the length of chains in $\Glog$.

It is for this reason that it is important to (at least partially) decouple the logical block order of $\Glog$ and the data dependency order $\Gsch$ which the decoding of logical blocks should respect.
The reason we say \textit{partially decouple} here is because, feed-forward classical control logic in the quantum circuit will need to be respected by both the decoder scheduling $\Gsch$ as well as consistent with the logical block order.
Each logical membrane associated to a logical outcome must be fully decoded before said outcome can be used to condition forthcoming circuit elements.
An extreme case of this is presented in Fig. \ref{fig:SR_rotation1}, where each $T$ gate implemented via magic state injection leads to a logical outcome which is needed to complete the consumption of the following magic state. 

\begin{figure}[ht!]
    \centering
    \includegraphics[width=0.95\linewidth]{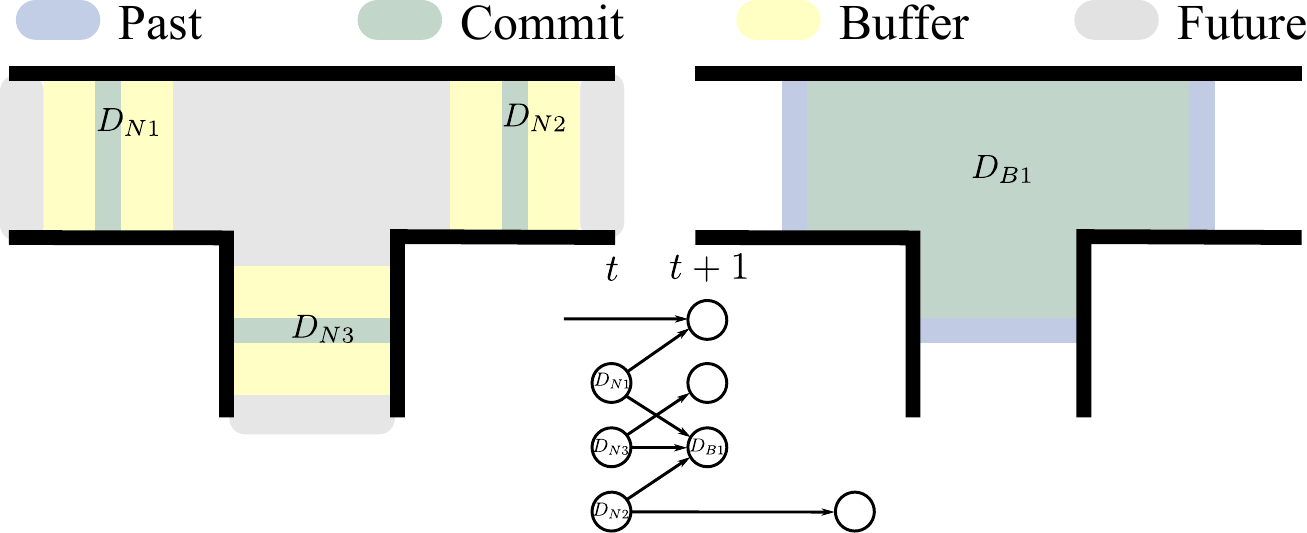}
    \caption{
    The figure depicts a four decoding tasks which provide an example of edge - vertex decoding.
    \textbf{(left)} Three edge decoding tasks $D_{N1}$, $D_{N2}$, $D_{N3}$ are solved independent of any other decoding task.
    The outcome data for a significant buffer regions (yellow) must be provided to these tasks in order for them to provide high quality recovery estimates on the commit regions (green).
    Edge decoding tasks treat unavailable outcomes (grey) as open boundary conditions.
    Such outcomes may be unavailable, either because they are yet to be produced, or because they are too far from the commit region to be relevant.
   \textbf{ (right)} A vertex decoding task is solved (green) with check boundary conditions (blue) imposed by preceding edge decoding tasks.
    \textbf{(bottom center)} The corresponding piece of the scheduling graph $\Gsch$, the DAG associated to the decoding tasks presented in the figure.
    Edge decoding task rely directly on measurement outcomes but not on corrections obtained by any other decoding task.
    In surface code lattice surgery, each edge decoding task provides boundary conditions to two vertex decoding tasks.
}
    \label{fig:EdgeVertexDecoding}
\end{figure}

\textbf{Parallel vertex decoding.}
An approach that gives a faster reaction time is to consider a partitioning of the vertices of $\Gsch$ into a graph coloring, where neighboring vertices have different colors. 
Indexing the colors by integers $\{1,2,\ldots , n_c\}$, one can perform all available tasks of the same color in parallel, starting with $1$ and proceeding in order.
We will still maintain one edge of $\Gsch$ per edge of $\Glog$, however, their direction will follow increasing color labels.
This gives a bound on the reaction time which is proportional to the number of colors ($n_c$). 
In particular, if $\Glog$ is bi-colorable, we can use this coloring to construct $\Gsch$ with a very low reaction time, since all causal chains will have depth bounded by $2$ (decoding tasks). 
If sufficiently many decoder modules are available (i.e. decoding throughput is met), this guarantees a reaction time which will not increase with the length of logical dependency chains in $\Glog$ as would be the case with sequential decoding.
However, each decoding problem while still having volume of order $\mathcal{O}(d^3)$ may still be large, owing to the large combined buffer and commit sizes. 
We improve on this with \textit{edge-vertex} decoding.

\subsection{Edge-vertex decoding}

Here we consider schedules with even lower reaction time than parallel vertex decoding. 
The approach works by first decoding ports between logical blocks, using the neighbouring blocks as buffers.
The commit region associated to ports is chosen to be a minimum number of error generator ``layers'' such that the Tanner graph for the decoding problem becomes disconnected in the direction perpendicular to the port, if the corresponding vertices are removed.
This effectively decouples the decoding problem for neighboring blocks.
After the ports are committed to, the blocks themselves may be decoded, using no extra buffers as their boundary conditions are now fixed. 

\textbf{Parallel edge-vertex decoding.}
The scheduling graph $\Gsch$ is constructed as follows. For each vertex and edge of $\Glog$ (corresponding to logical blocks and ports, respectively), we place a vertex for $\Gsch$. 
We place a directed edge between a block vertex and any port vertex that belongs to it, directed from the port vertex to the block vertex. 
Buffers for the port vertices are given by the neighbouring logical blocks. The name of the scheme is derived from the fact that edges of the logical block network are decoded first, followed by the vertices. We give examples of this schedule in the following section.
In terms of reaction time, the scheduling graph $\Gsch$ is bipartite (i.e., depth of $2$).
This scheme thus has an extremely low reaction time, and as we will numerically be show to performs extremely well in terms of logical error rate.

\section{Simulation and results \label{sec:simulations}}

\begin{figure}[ht!]
    \centering
    \includegraphics[width=0.95\linewidth]{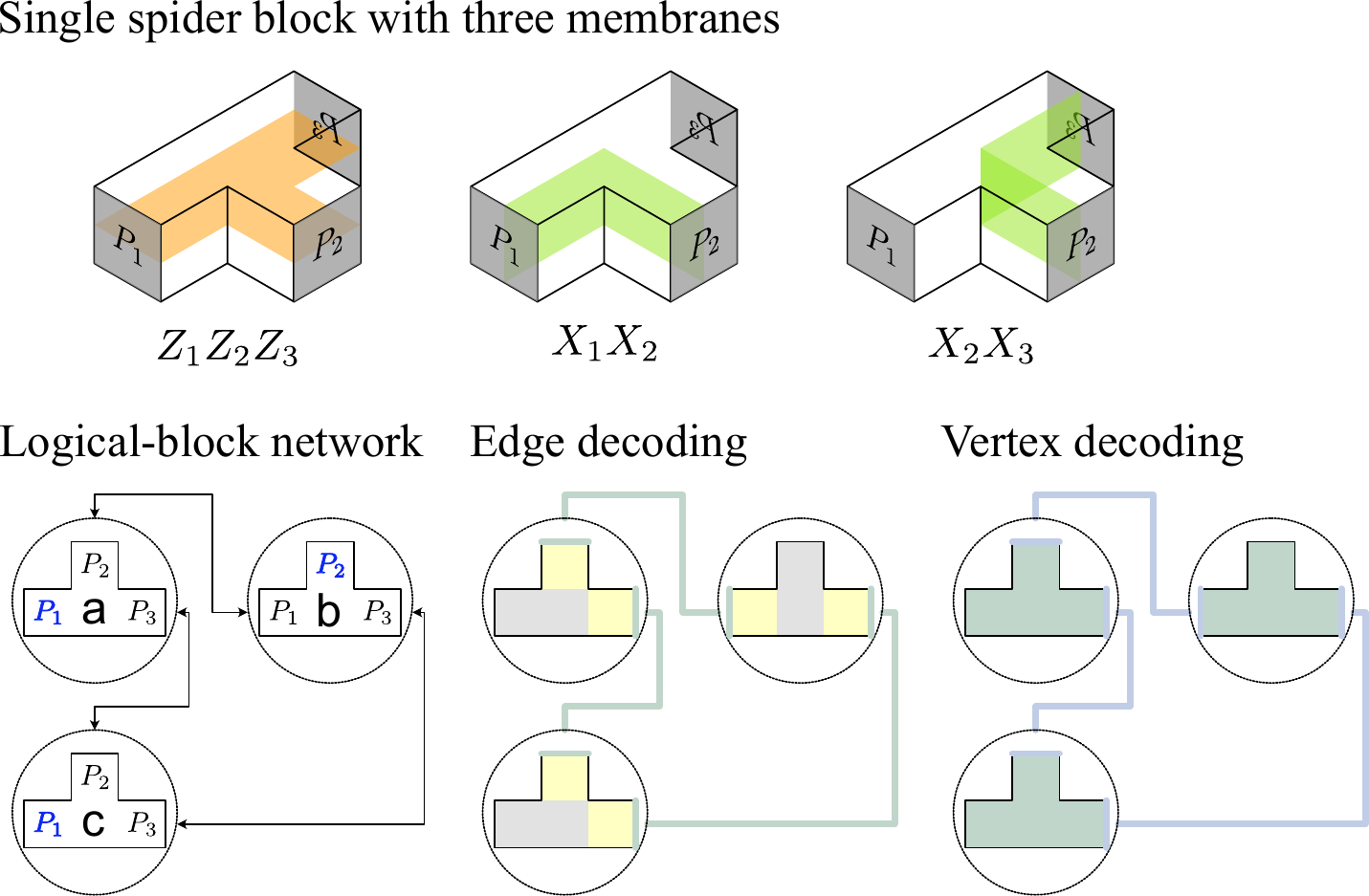}
    \caption{Schematics of the software implementation of modular decoding. 
    \textbf{(top)} An elementary logical block with three ports and three membranes, where each membrane is labeled with the corresponding stabilizer representation. 
    \textbf{(bottom left)} A simple logical block network with three vertices $\{a, b, c\}$ and three edges $\{(a,P_2)\leftrightarrow(b,P_1),(a,P_3)\leftrightarrow(c,P_2),(b,P_3)\leftrightarrow(c,P_3)\}$. 
    Here each edge represents an identification of two matching ports from two blocks.
    Global logical membranes are expressed in terms of stabilizers on the open ports (blue), e.g., there is a global membrane $Z_{a,P_1}Z_{b,P_2}Z_{c,P_1}$, which is the union of all three individual $ZZZ$-type membranes. 
    \textbf{(bottom right)} edge-vertex modular decoding. 
    The interface regions (or \emph{edges} in the logical block network, shown in green) are decoded first in parallel, with buffers growing into bulk regions (yellow). 
    After the edge decoding, remaining disconnected bulk regions (or \emph{vertices} in the logical block network) are decoded in parallel. 
    The edge decoding step sets boundary conditions (blue) for the vertex decoding step.}
    \label{fig:schematics}
\end{figure}

In this section, we first provide a short description of our implementation of the modular decoding scheme for logical block networks, then present extensive simulation results for logical block networks with increasing complexity. 
We start from a benchmark model of a linear chain of identity blocks.
We then move to a planar network of ZX-instruments with a more complicated composition of logical membranes.
Finally, we simulate the Clifford part of the 15-to-1 magic state distillation protocol. 
In each case, we show that with sufficient buffering, the performance of modular decoding approaches that of monolithic decoding.

\subsection{Software implementation}

We describe the modular decoder implementation used in our numerical investigation. 
Firstly, we consider the implementation of the logical block networks. 
Each elementary logical block defines its own labeled input and output ports.
For example, the identity block has an input port and an output port named "IN" and "OUT" respectively. 
The partial membranes in each logical block are labeled by their stabilizer representations, e.g., the two partial membranes of the identity block are labeled $X_\mathrm{IN}X_\mathrm{OUT}$ and $Z_\mathrm{IN}Z_\mathrm{OUT}$. 
Then one can define a logical block network by adding elementary logical blocks as nodes in the network, and specifying the matching of ports among the blocks as edges in the network. 
The buffer regions (after specifying the buffer size), global membranes, and their specific decomposition into partial membranes are then computed automatically. 
As an example, we can define a chain of two identity blocks as follows: Add two identity blocks as vertices $a$ and $b$ into the network; add an edge $e = (a,\mathrm{OUT}) \leftrightarrow (b,\mathrm{IN})$. 
The buffer region will grow (by syndrome-graph traversal) around $e$ until reaching the target buffer size. 
Global membrane decomposition will be derived, e.g., there will be a global membrane $X_{a,\mathrm{IN}}X_{b,\mathrm{OUT}}$, which is the union of two local membranes represented by $X_{a,\mathrm{IN}}X_{a,\mathrm{OUT}}$ and $X_{b,\mathrm{IN}}X_{b,\mathrm{OUT}}$. 
This simple design applies to arbitrary logical block networks. 
See Fig.~\ref{fig:schematics} for schematics for a slightly more complicated example.
For the purposes of simulation, any remaining input and output ports of the global problem are treated as a noiseless readout. (See Sec.~XII. M. of Ref.~\cite{bombin2021logical} for more details on this approach.)

\subsection{Noise model}
For the simulations, we assume that the fault-tolerant logical blocks are realised using fusion-based quantum computation with the $6$-ring fusion network~\cite{bartolucci2021fusion, bombin2021logical}. Fusions, which are bell basis measurements $\{XX, ZZ\}$ are performed between resource states, which are ring-like cluster states on $6$ qubits. Boundaries are realized using certain single qubit measurement patterns. Checks are constructed out of outcomes of fusion measurements: those in the bulk consist of $12$ fusion outcomes, while those on the boundary (involving single qubit measurement outcomes) may involve $8$ or fewer outcomes. We use the \textit{hardware-agnostic error model} of Ref.~\cite{bombin2021logical}, where each fusion outcome and single-qubit measurement outcome is subject to an \textit{i.i.d} bit-flip error with probability $p_{\mathrm{error}}$. Under this error model, using the Union-Find decoder, the threshold error rate is $p_{\mathrm{error}}^*=0.95\%$. We will look at fixing the error rates at $p_{\mathrm{error}}=0.5\%$, which is approximately half the threshold.

We remark that despite the numerical results being based on fusion-based quantum computation, we expect the qualitative results and conclusions to extend to circuit-based and measurement-based implementations of surface code computations. This is because the decoding problem for all approaches can be expressed in terms of a syndrome-graph, and one may regard the differences as a choice of error model. See secs. VII. and VIII. of Ref.~\cite{bombin2021logical} for more details on the scheme and error model.

\subsection{Idling memory}

The simplest model for testing modular decoding schemes is in the case of a logical qubit in memory, which can also be thought of as a logical identity block (or chain thereof). 
The functionality of the full network, in this case a chain of $\tau$ identity blocks, is still the identity gate and the two global logical membranes $X_0 X_\tau$ and $Z_0 Z_\tau$ are simply concatenations of the two local logical membranes $X_t X_{t+1}$ and $Z_tZ_{t+1}$ for $t\in\{0,\ldots,\tau-1\}$. 
Nevertheless, this setup is enough to show important aspects of modular decoding, including the need for buffering and optimization of scheduling.

\begin{figure*}[ht!]
    \centering
    \includegraphics[height=76mm]{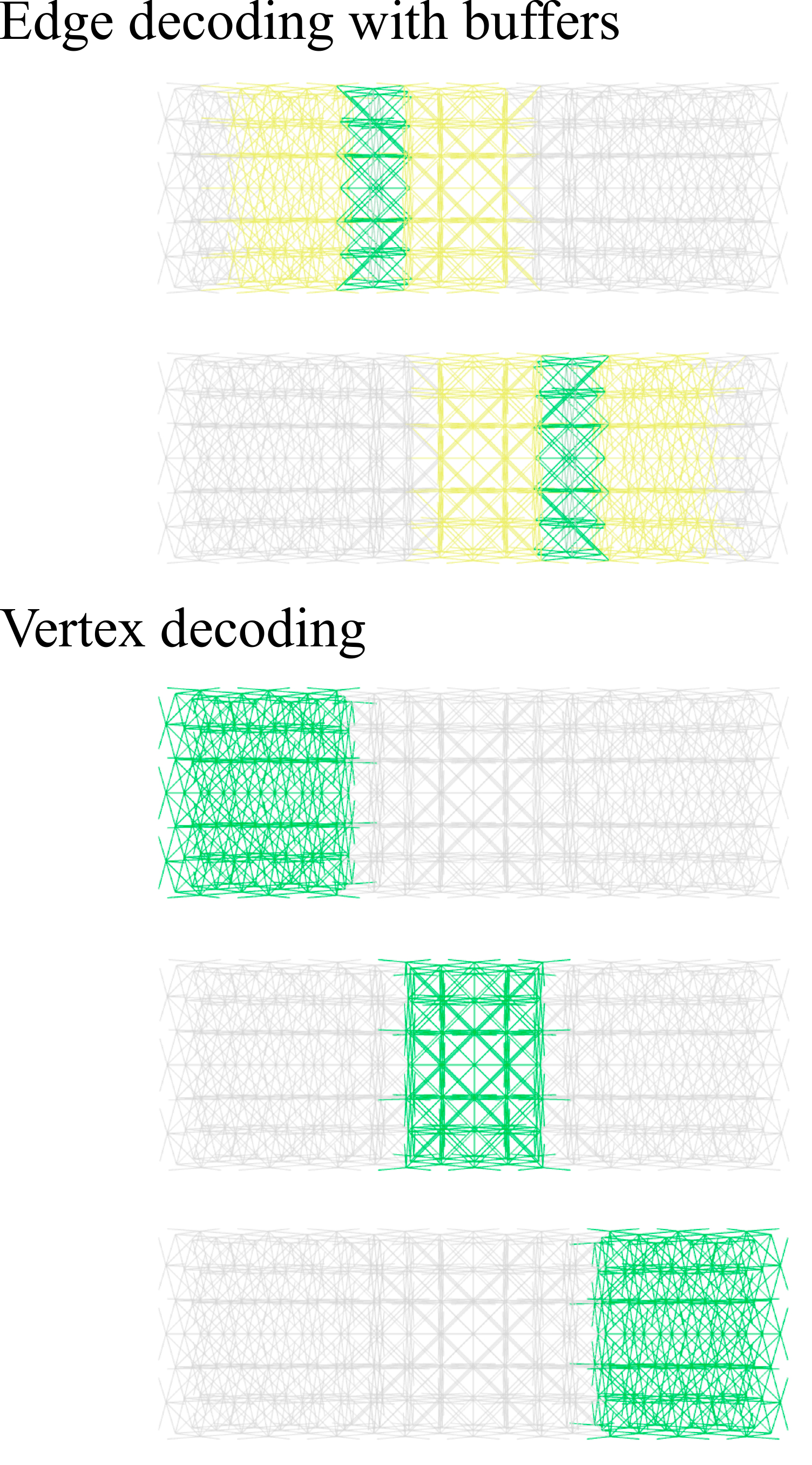}
    \quad
    \includegraphics[height=80mm]{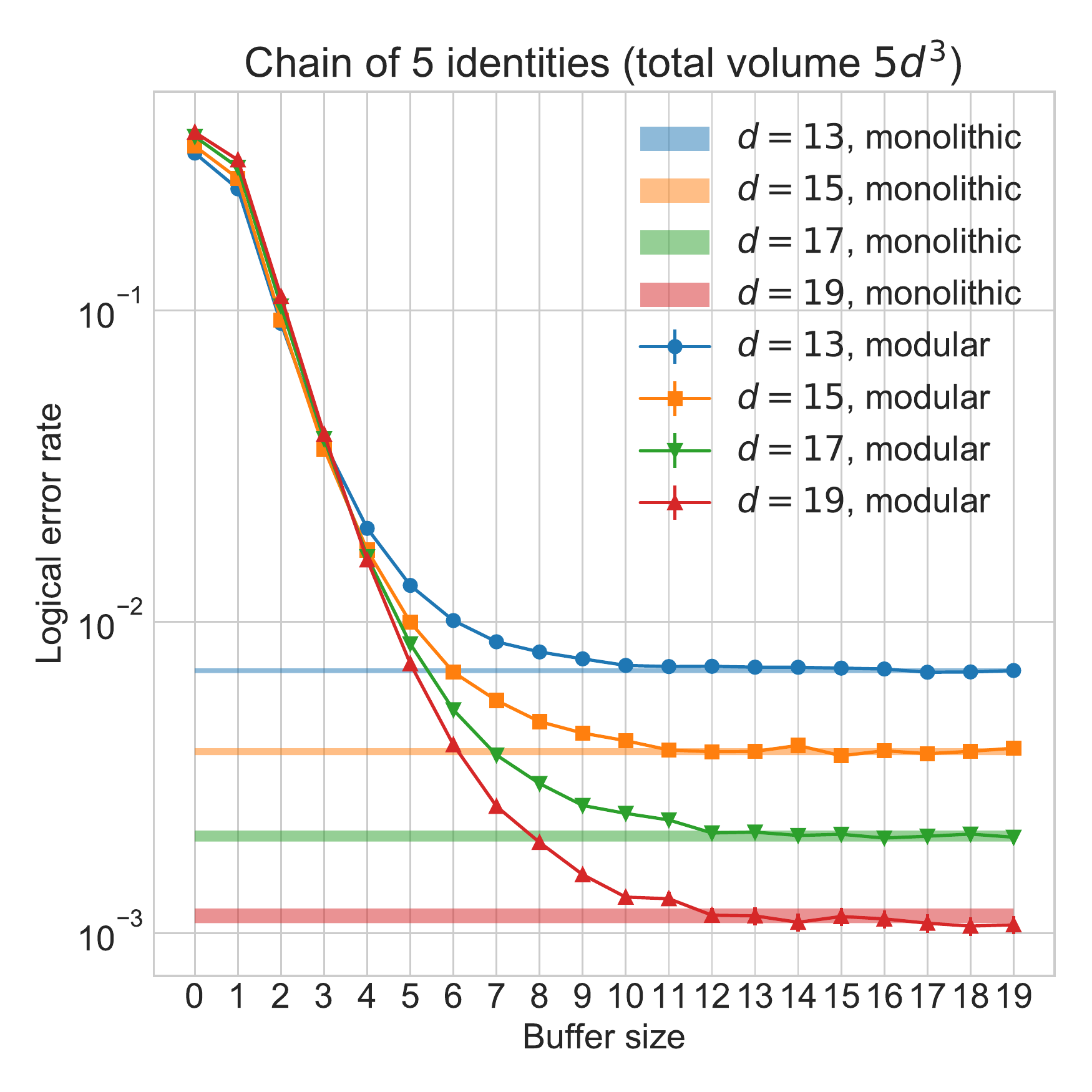}
    \caption{Modular decoding for a chain of identity blocks. 
    \textbf{(left)} Schematics of edge-vertex modular decoding for a chain of 3 identity blocks. In the first edge-decoding step, all interface regions (green) are decoded in parallel with buffers (yellow) grown into neighbouring bulk regions; in the second vertex-decoding step, the bulk regions (green) are decoded in parallel. Here we show the primal syndrome graph, and the LER corresponds to that of the membrane $X_0X_5$. Decoding of the dual syndrome graph is similar. 
    \textbf{(right)} Impact of buffer size $b$ on modular-decoding logical error rate (LER) for a variety of protocol fault distances $d$ (simulation performed for a chain of 5 identity blocks). 
    Without buffering, the LER is close to 50\% per logical membrane; increasing $b$ quickly decreases the LER until it becomes indistinguishable from that of monolithic decoding. 
    Note that in this case the two global logical membranes ($X_0X_5$ and $Z_0Z_5$) have the same logical error rate.}
    \label{fig:identity_chain}
\end{figure*}

We compare the LER among monolithic decoding (using the union-find decoder), and modular decoding with different buffer sizes (where each modular decoder also utilizes union find). 
Fig.~\ref{fig:identity_chain} shows the evolution of LER with increasing buffer sizes, for a variety of protocol fault distances $d=13, 15, 17, 19$. 
Without buffering (buffer size $b=0$), the LER is close to 50\%. 
The LER decreases exponential with initial increase of the buffer size $b$.
However, as the buffer size $b$ approaches the protocol fault distance $d$, this improvement in LER stagnates.
For $b\geq d$, the buffering condition of Def.~\ref{def:buffering_condition} is satisfied and numerically obtained LER from modular decoding is indistinguishable from monolithic decoding.

\subsection{Planar network of ZX-instruments}

\begin{figure*}[ht!]
    \centering
    \raisebox{-0.5\height}{\includegraphics[width=0.27\linewidth]{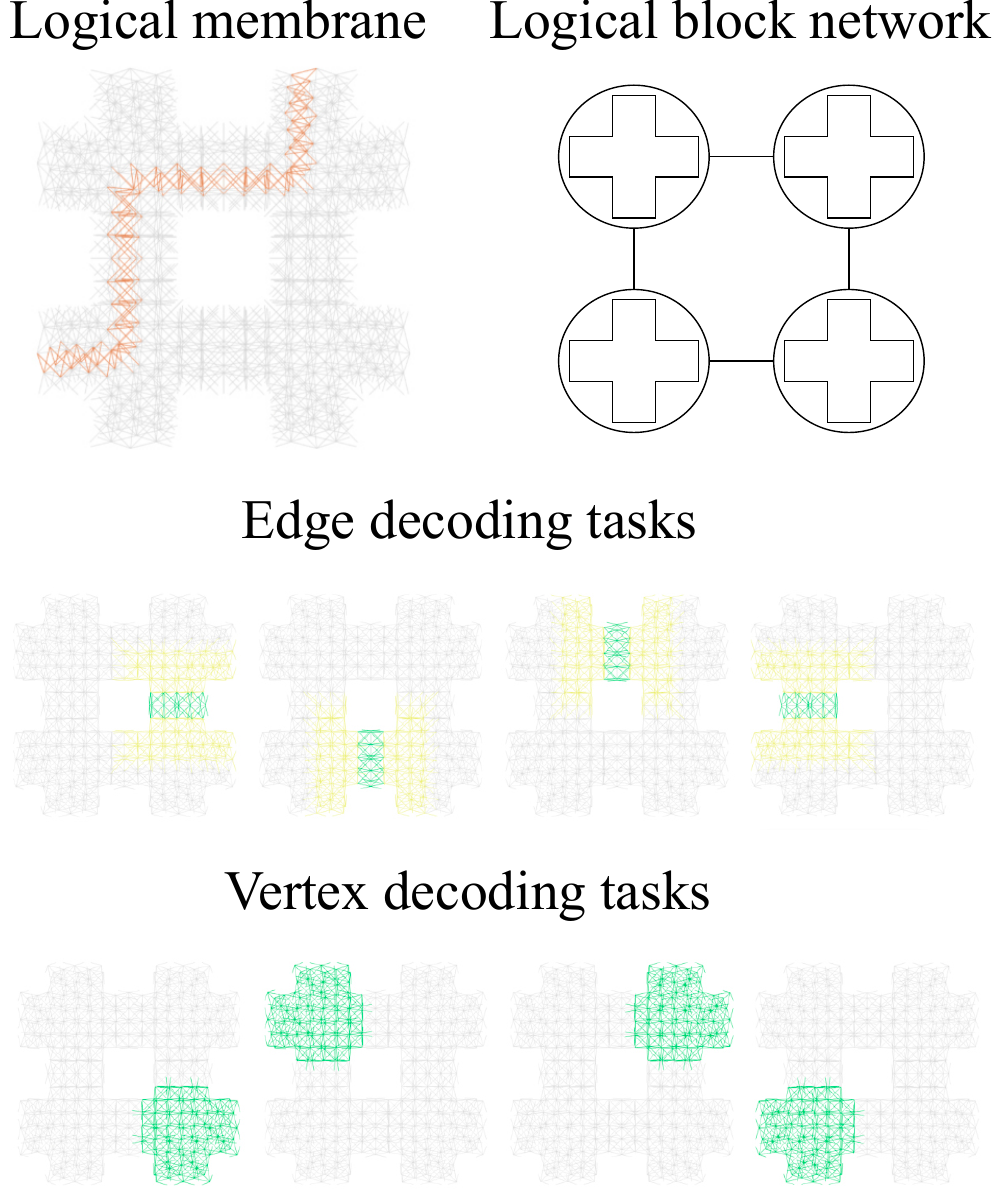}}
    \raisebox{-0.5\height}{\includegraphics[width=0.71\linewidth]{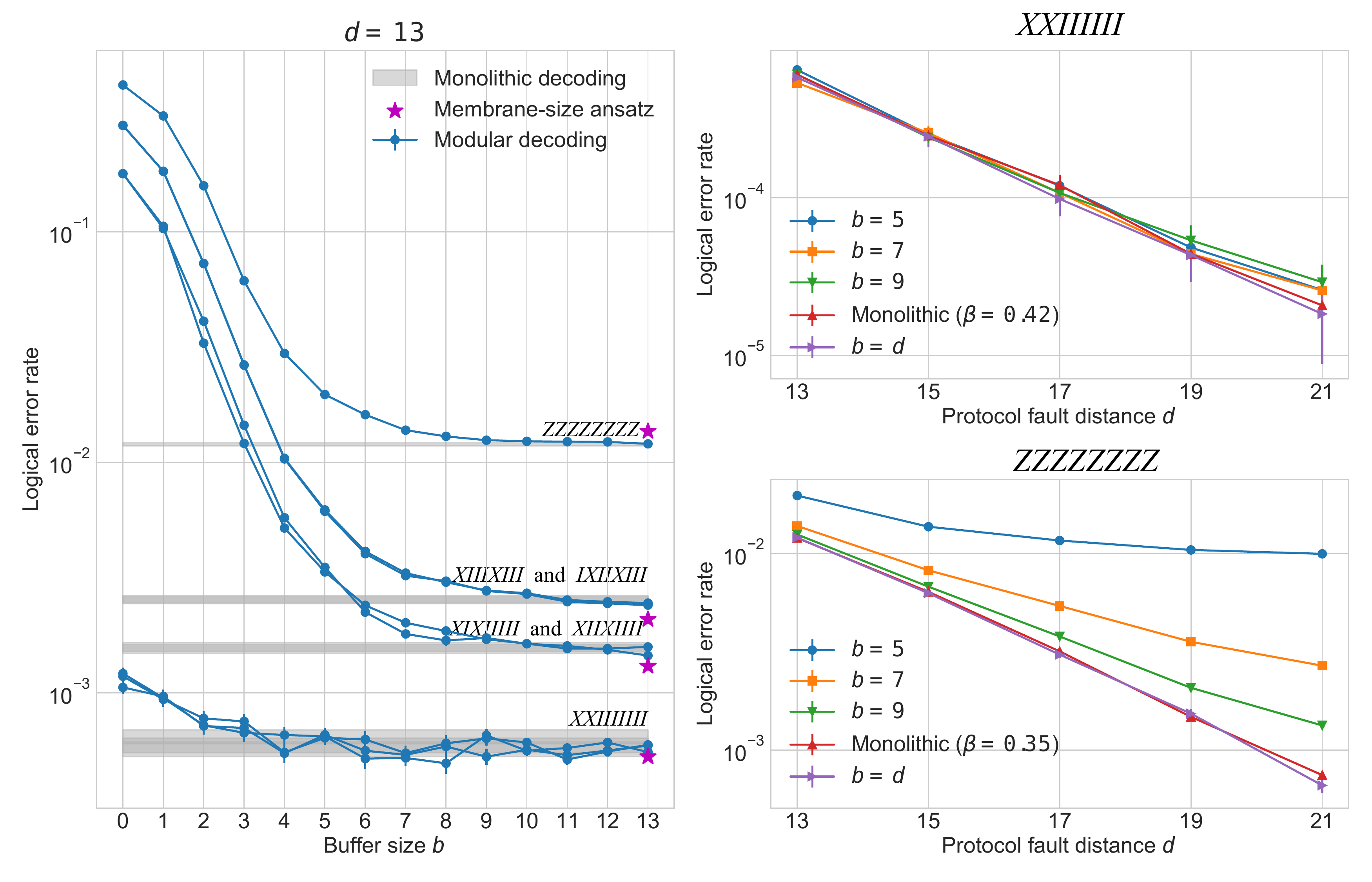}}
    \caption{Edge-vertex decoding on a planar network of logical blocks. 
    \textbf{(left)} Logical block network and syndrome graphs for the protocol. 
    The network consists of four 4GHZ logical blocks connected in a ring structure to form an 8GHZ with one meta-check. 
    Orange shading shows one of the logical membranes (corresponding to $IIXIIIIX$ stabilizer), supported on three of the constituent blocks and crossing two port connections. 
    Commit and buffer regions are respectively shaded green and yellow for all decoding tasks (both edge and vertex).
    \textbf{(middle)} Impact of buffer size $b$ on logical error rate (LER) in edge-vertex modular decoding. 
    Each curve corresponds to a different logical membranes in the network and are labeled up to a cyclic shift by 2 and order inversion (i.e. the geometric symmetries of the network).
    Membranes crossing a larger number of ports have higher LER for low $b$.
    For sufficiently large buffer size $b$, all LERs approach to values from monolithic decoding (shaded horizontal gray lines), which is roughly proportional to membrane size (purple star estimate). 
    \textbf{(right)} Logical decay $\beta$ for selected logical membranes (using stabilizer representation) and buffer sizes $b$. 
    The $XXIIIIII$ logical membrane is supported on a single logical block and is minimally affected by buffer size (beyond $b\geq5$).
    The $ZZZZZZZZ$ logical membrane is supported on all four logical blocks and all four ports. It is maximally sensitive to buffer size $b$ and increasing code distance $d$ does not yield exponential suppression of LER for $d\gg b$.}
    \label{fig:spider_ring_data}
\end{figure*}

In this example, we study a planar network of four logical blocks (see Fig.~\ref{fig:spider_ring_data}). 
Each logical block supports partial membranes $Z_1Z_2Z_3Z_4$ and pairwise $X_iX_j$'s (i.e. a 4GHZ entanglement structure). 
The logical block network presented has 8 ports, whose partial membranes are $\prod_{i=1}^8Z_i$ as well as pairwise $X_iX_j$'s (i.e. a 8GHZ entanglement structure).

In addition to the logical membranes which provide the logical stabilizers for this fragment there is a logical \textbf{meta-check}.
This meta-check is beyond the scope of the topological fault-tolerance, to which modular decoding is being applied.
Meta-checks are the basis for concatenating fault-tolerant protocols and are not included in the group $\checkgrp$ generated by the local check generators $\checkgen$.
In this example, the meta-check corresponds to a closed membrane composed of four partial $XX$ membranes on the constituent logical blocks but with no support on external ports.
In the case of correct topological decoding the "logical" outcome associated to this membrane yields a fixed value.
Obtaining a different value is indicative of an error promoted to a logical error by topological decoding, which can nevertheless be caught by the meta-check.

Fig.~\ref{fig:spider_ring_data} provides numerical data confirming that the LER from modular decoding approaches that from monolithic decoding quickly with increasing buffer size $b$.

A key feature of this logical block network is that global membranes come in many different sizes. 
Note that logical membranes are topological objects, and deformation of a logical membrane by applying (XOR with) parity checks on the syndrome graph simply leads to an equivalent logical membrane. 
To quantify the size of a logical membrane in a topological way, we count the number of minimal-weight logical errors for the given membrane (i.e., the minimal-weight undetectable errors that intersect the given membrane an odd number of times). This turns out to be a good proxy of LER especially at low physical error rate. We show that the LER is indeed roughly proportional to the size of the logical membrane. 
Specifically, we collect the LERs from all membranes $\{\mathrm{LER}(\mathcal{M}_i)\}$ and calculate a unit of LER, $p=\sum_i\mathrm{LER}(\mathcal{M}_i)/\sum_i\mathrm{size}(\mathcal{M}_i)$. The membrane-size ansatz reads: $\mathrm{LER}^\mathrm{ansatz}(\mathcal{M}_i)=p\times\mathrm{size}(\mathcal{M}_i)$.

The logical decay is useful for quantifying the FT capability of a protocol. It is the speed of the exponential decay of the LER with increasing block size $L$ when the physical error rate is below threshold, i.e., the $\beta$ obtained when fitting the logical error rate to the relation $\mathrm{LER} = \alpha e^{-\beta L}$ (where $\alpha$ is another fit parameter). 
A good FT protocol has large $\beta$, such that the target LER can be achieved by small $L$. 
We observe that with modular decoding, the logical decay increases with buffer size $b$, approaching to that of monolithic decoding; also, the logical decay on larger membranes are generally smaller than that on smaller membranes.

\subsection{15-to-1 magic state distillation protocol}

The magic state distillation (MSD) protocol is essential for universal quantum computing. 
We present the first fault-tolerance simulation of the \textit{static}, \textit{Clifford} part of the 15-to-1 MSD protocol. 
Here \textit{static} means we focus on the MSD protocol before the adaptive measurements of input magic states, and \textit{Clifford} means we ignore errors from the injection of input magic states and focus on errors incurred by the Clifford quantum gates. 
This part of the 15-to-1 protocol can be realized by a network of logical blocks using the tri-orthogonal matrix representation (which will be explored in an upcoming paper~\cite{MSDinprep}. 
Specifically, there are 27 logical blocks (16 red ones and 11 green ones), 27 global ports, and 27 global membranes with varying sizes. 
In the edge-vertex modular decoding scheme, 46 edge-decoding tasks will be performed in parallel first, followed by 27 vertex-decoding tasks. 

\begin{figure*}[ht!]
    \centering
    \raisebox{-0.5\height}{\includegraphics[width=0.35\linewidth]{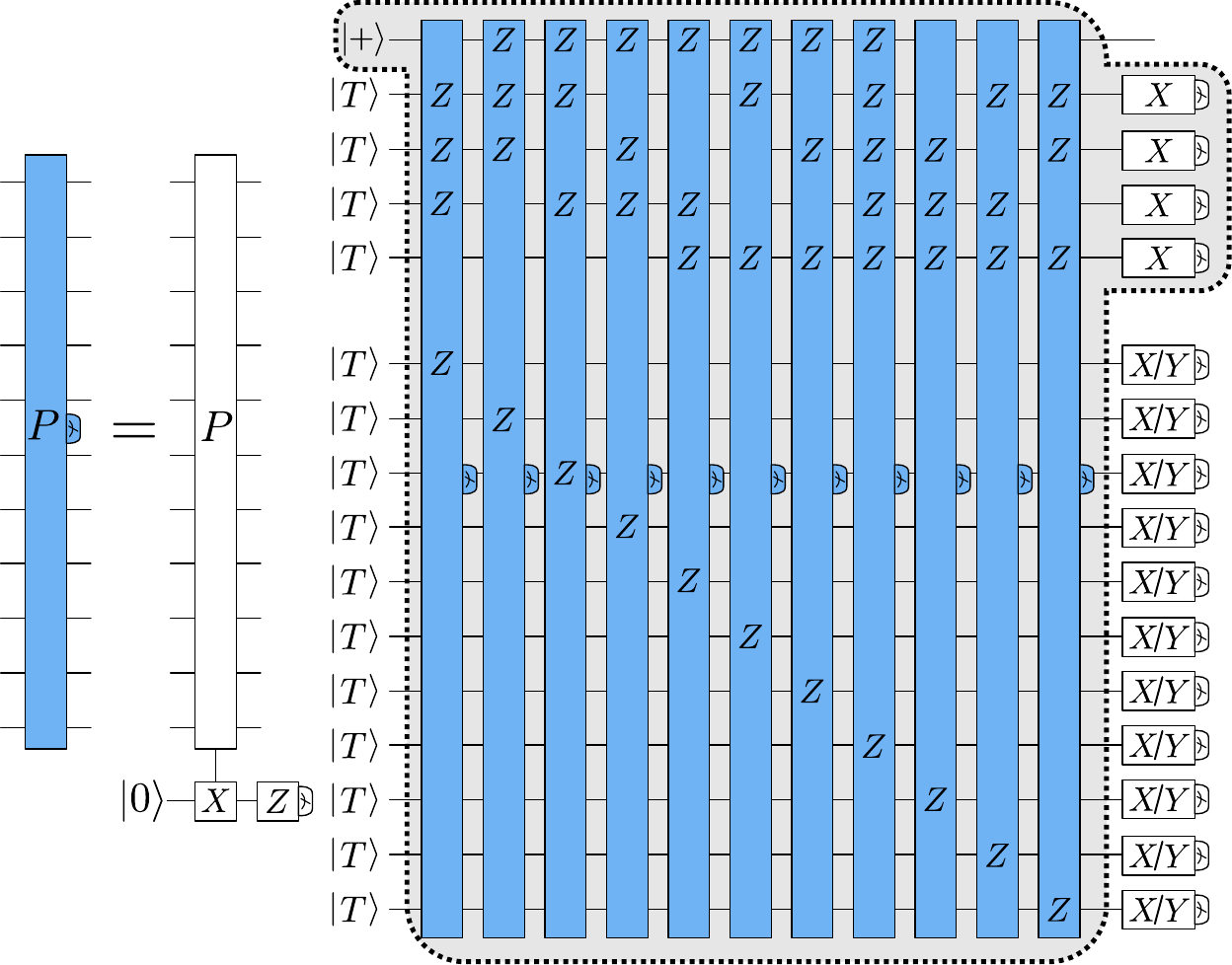}}
    \raisebox{-0.5\height}{\includegraphics[width=0.64\linewidth]{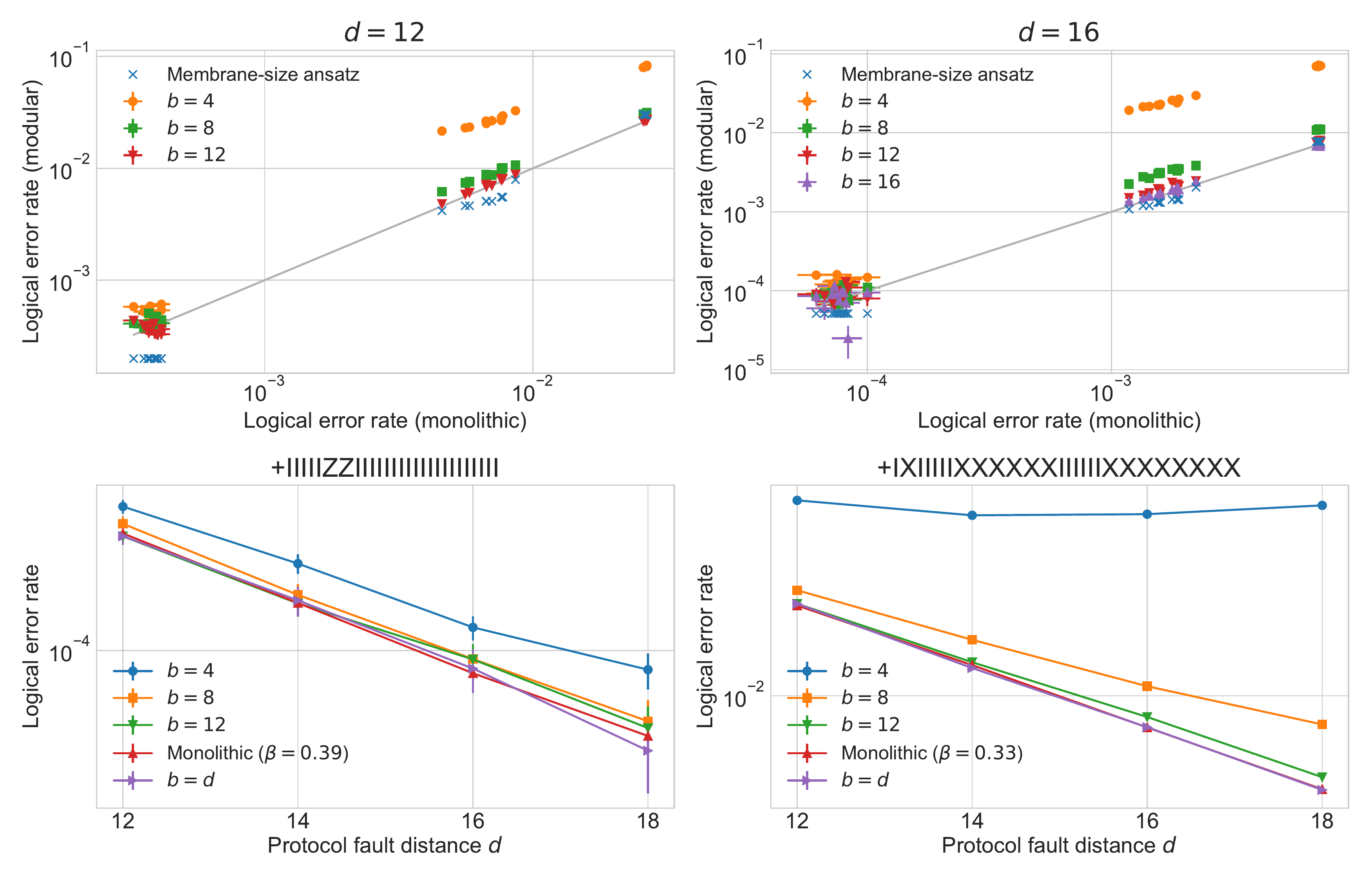}}
    \caption{
    (left)  A blue box notation for non-destructive Pauli measurement of $P$ is introduced to represent the quantum circuit for the 15-to-1 magic state distillation using 11 auto-corrected $T$ gates.
    The \textit{static Clifford} part is marked with a dashed outline and grey background.
    The circuit can be reduced to GHZ type logical blocks as in Fig.~\ref{fig:LogicBlocks} where modular decoding is performed in two steps.
    First, parallel edge decoding for all 46 interfaces with buffers, followed by parallel vertex decoding for all the 27 interiors of the GHZ style logical blocks.
    (right) LER performance of modular decoding on a static Clifford portion of a 15-to-1 magic state distillation protocol. 
    (upper panels) Logical error rate (LER) from modular decoding versus that from monolithic decoding, for the 27 logical membranes (each of the 27 points of a given color represents a logical membrane), and different buffer sizes. 
    With increasing buffering, the performance of modular decoding approaches to monolithic decoding (indicated by the gray $y=x$ line). 
    Larger membranes have larger LERs, and fitting of the LER according to the size of the corresponding logical membrane (blue crosses) agrees well with the actual data. 
    Here the physical error rate is 0.005, which is around half of the fault-tolerance threshold. 
    (lower panels) The left/right panel shows the logical decay for a representative small/larger membrane, which has higher/lower decay constant $\beta$, and is less/more sensitive to the buffer size.}
    \label{fig:msd_data}
\end{figure*}

Fig.~\ref{fig:msd_data} shows edge-vertex modular decoding with buffering still performs very well for this fairly large logical block network. 
In this example, different logical membranes have LERs that are differed by orders of magnitude, because of the drastic difference in the membrane sizes. 
The LER is still roughly proportional to the membrane size. 
The logical decay rate $\beta$ still increases with buffer size, and is in general smaller on larger membranes.

\section{Conclusions and outlook}\label{sec:Conclusions_and_outlook}

We have introduced {\it modular decoding}, a distributed approach to solving large decoding problems by decomposing them into smaller decoding sub-tasks with minimal data inter-dependencies.
Each sub-task can be solved by a suitable off-the-shelf, offline decoder (such as MWPM~\cite{dennis2002topological,kolmogorov2009blossom} or UF\cite{delfosse2017almost} for surface-code based schemes) to produce a partial recovery, which are then combined to produce a global correction. 
Modular decoding is designed to guarantee the availability of logical measurement outcomes as they become necessary for branching decisions throughout the quantum computation. 
In order to provide these outcomes in a timely manner and avoid slowing down the computation, the decoder must keep up with the continuous stream of measurement data produced. 
This places stringent requirements on the the decoder subsystem, such as high throughput (keep up with the overall data rate), short reaction times (provide partial results within a short time frame) and minimal LER (logical error rate).

The main novel features of our decoder, are scalable throughput and low reaction time and these are achieved by design.
Scalable throughput is guaranteed by the high degree of parallelism which is achieved by distributing independent decoding tasks to different decoder modules which operate synchronously, as per Fig.~\ref{fig:sys_overview}.
Low reaction time (or at least the computational contribution to it) is achieved by making decoding sub-tasks relatively small (i.e. $O(d^3)$ error and check generators, for fault-distance $d$) and by keeping all data dependence chains in the scheduling graph $\Gsch$ short.
The most extreme example for this is given by what we refer to as \textit{edge-vertex} decoder scheduling which minimizes both task size and data dependence chains and derives its scheduling graph from the connectivity graph $\Glog$ of the logical block network.

While the main novel features of our approach are essentially guaranteed by construction, most of our work goes into understanding how to retain a competitively low logical error rate (LER) w.r.t. existing decoders.
To this end, we provide examples, a rigorous soundness proof (Sec. \ref{sec:soundness-proof}) and numerical evidence (Sec. \ref{sec:simulations}).
All of these lead us to the same consistent conclusion;
In order to retain the original fault-distance of the protocol as well as similarly low LER, it is necessary to supplement each decoding task with a buffer of syndrome information along a neighborhood of width roughly $d$ (the protocol distance).
This is one of the assumed conditions in proving decoder soundness and is found to be necessary and sufficient to achieve an error rate indistinguishable from that of a monolithic decoder with simultaneous  access to all syndrome information.

While union find decoder \cite{Delfosse_2021} already have an almost linear complexity in the size of the decoding problem input, there are other decoders such as minimum-weight perfect matching \cite{fowler2013minimum} or tensor network decoders \cite{bravyi2014efficient, darmawan2017tensor, Darmawan_2018} which have a significantly less favorable scaling.
While our numerical simulations used a union-find decoder as a base decoder for individual tasks in modular decoding, the soundness proof for modular decoding is completely general.
The modular decoding decomposition provided by edge-vertex decoding or other variants guarantee that the computational scaling of the base decoder will only be relevant up to inputs of size $O(d^3)$ beyond which, a linear coarse grained scaling guaranteed by modular decoding kicks is.
As such, we expect modular decoding provides a way to \textit{linearize} the coarse grained complexity of arbitrary decoding algorithms without compromising their decoding accuracy.
This allows seriously considering other computationally costly decoding algorithms with higher noise thresholds and use them as a base decoders for modular decoding. 

The methods, and scheduling sections (\ref{sec:MDmethods} \& \ref{sec:scheduling}), accurately describe how to decompose a global decoding problem into sub-tasks, schedule these and identify the necessary buffer regions for each task (Sec. \ref{sec:buffer_growth}) in a way which satisfies all of the desired conditions.
Our prescriptions are most concrete for topological quantum circuits (also known as lattice surgery), for which the decomposition into decoding tasks mirrors the graph structure $\Glog$ of the logical block network.
The simple but powerful buffering method presented (Sec. \ref{sec:buffer_growth}) is, to our knowledge, a novel contribution.
When used with buffer parameter $b=d$ (i.e., the fault-distance of the protocol), the modular decoding scheme obtained is guaranteed to satisfy the decoder soundness.
More importantly, the buffering method, is put to the test in combination with edge-vertex decomposition approach and is numerically shown to perform extremely well under the same condition.
Moreover, we show the robustness and flexibility of this method by decoding a Clifford circuit fragment of 15-to-1 magic state distillation which is composed of 27 logical blocks with 46 connections among them.

In summary, we have defined modular decoding and shown that it can be instantiated to meet the practical requirements associated with real-time decoding: high throughput, short reaction time and low LER.
We look forward to its hardware implementation supporting real-world fault-tolerant quantum computations.

\textbf{Future directions.}
While our approach minimizes the impact of the decoding process on the reaction time, 
in practice, there are further hardware and systems considerations that are relevant.
Improvements to the speed of individual decoder units (also known as offline decoders) remain important directions. 
One can combine our modular approach with the complementary approaches of \textit{pre-decoding} and data compression to further reduce the reaction time. 
In these approaches, the decoder accuracy is (slightly) sacrificed in order to simplify or speed up the decoding problem~\cite{delfosse2020hierarchical,das2020scalable,smith2022local}. 
The buffer size offers another tunable parameter to trade off reaction-time and logical-error rate performance. 

We remark that although we have focused on fault-tolerant schemes based on the surface code, our modular decoding schemes (such as edge-vertex decoding) can readily be applied to any universal computation for fault-tolerant computations based on topological error-correcting codes.
For example, our scheme can be applied to color codes in various dimensions~\cite{bombin2007exact, bombin2012universal, kubica2015unfolding, bombin2016dimensional, bombin20182d}, subsystem color codes~\cite{bombin2010topological, kargarian2010topological, roberts20203}, and floquet codes~\cite{hastings2021dynamically, paetznick2022performance}, and it would be interesting to see if the accuracy is maintained for similar sized buffers. 
A suitable decoder for sub-tasks will be required in each of these cases. 
Belief propagation with ordered statistics decoding (BP OSD)---which has shown to have quite good performance for a range of codes~\cite{panteleev2021degenerate, roffe2020decoding})---along with renormalization group decoding~\cite{duclos2013fault} are examples of suitable decoders.

Beyond topological codes, it may be interesting to apply modular decoding techniques to the setting of quantum LDPC codes~\cite{kovalev2013fault,breuckmann2021quantum,panteleev2022asymptotically}, or combine them with existing parallel decoders~\cite{leverrier2022parallel}. As quantum LDPC codes with good code-properties require high expansion~\cite{baspin2022connectivity}, care is needed to prevent the buffered sub-task from becoming too large. 

Finally, it would be interesting to extend the soundness theorem to prove a fault-tolerance threshold theorem for universal computation with surface codes (i.e., a topological analog of the threshold theorem for concatenated codes in Ref.~\cite{aliferis2005quantum}). In particular, one can readily adapt the argument for the lower bounding accuracy threshold in Ref.~\cite{dennis2002topological} to include the use of Buffers. In particular, applying this argument to the Clifford part of the CNOT architecture in Fig.~\ref{fig:LogicBlocks} can give a lower bound on the fault-tolerant threshold for logical Clifford operations. Taking the minimum of this, and the threshold for distilling magic states~\cite{bravyi2005universal}, one can obtain a threshold theorem for universal computation with surface codes.

\section{Contributions and acknowledgements}
CD and NN performed the first study in modularizing the decoding problem.
CD developed the decoding framework used for numerical simulations.
FP and SR performed early research on modular decoding identifying the need for buffering.
HB proved soundness of modular decoding under the buffering condition.
SR implemented the elementary logical block.
YL proposed and implemented the buffer growth algorithm, defined and implemented the logical block networks, implemented the modular decoding simulations and performed the numerical experiments presented in this article.
HB has not taken part in the writing of the article or reviewed its final form.
YL, FP, SR contributed to writing and reviewing the article as well as figure production.
We thank D. Litinski for numerous useful discussions, figures and feedback.
We thank all our colleagues at PsiQuantum for helpful discussions and feedback on the draft but especially Terry Rudolph, Dan Dries and Mercedes G. Segovia.
FP, YL, and SR would like to dedicate this article to the memory of David Poulin, an inspiring researcher and mentor who left us too soon.
\appendix

\section{Modular decoding system-level overview}\label{app:system_data_flow}
\begin{figure*}[h!]
\includegraphics[width=0.85\linewidth]{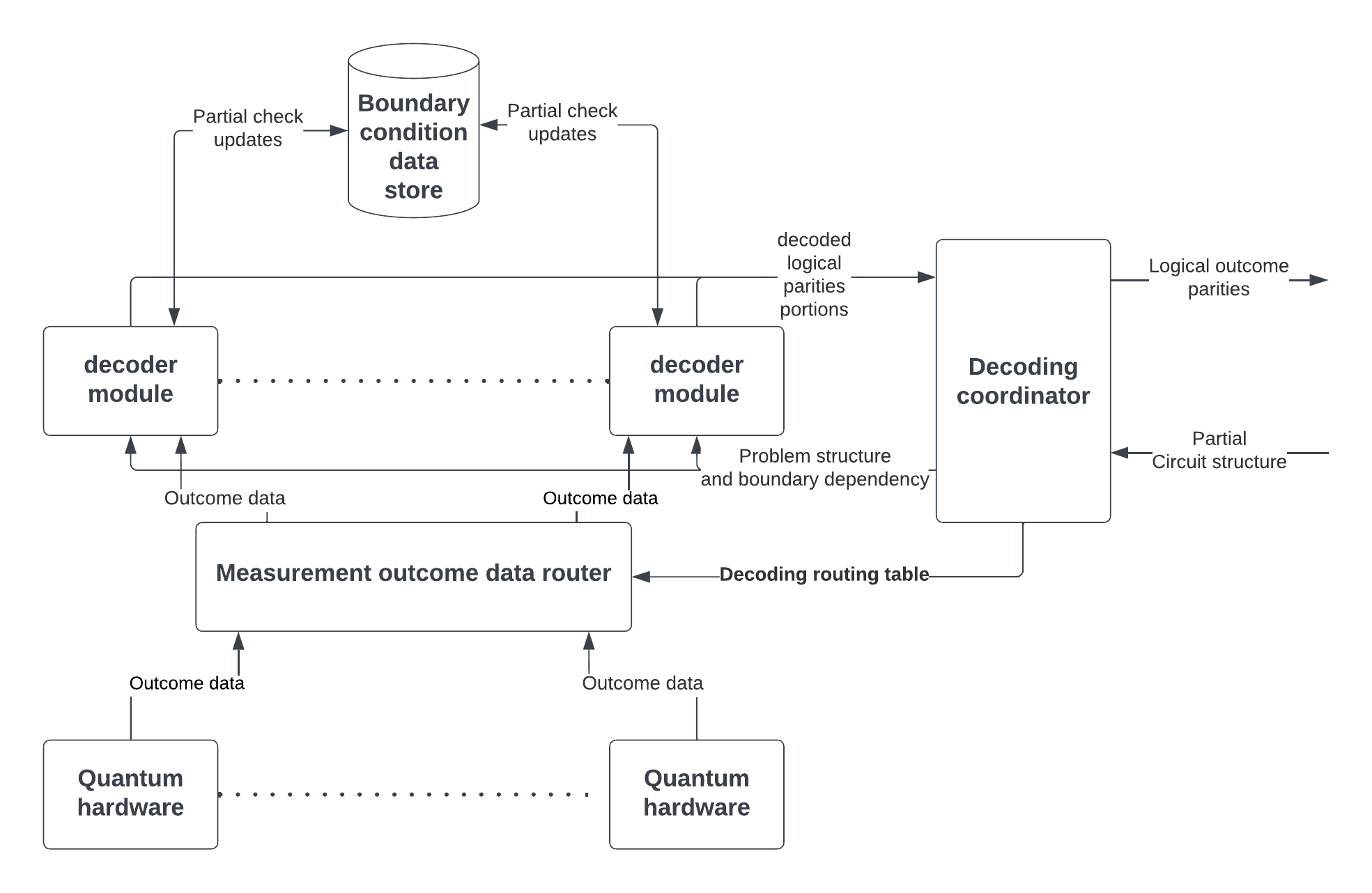}
\caption{A schematic for the system-level data flow in a modular decoding implementation. the decoding coordinator has a description of the logical block network, and partitions and schedules the global decoding problem into sub-decoding problems. The decoding coordinator also combines partial membrane outcomes into global membrane outcomes. The quantum hardware units produces a stream of classical outcomes (which may or may not be partially compressed). The measurement outcome data router receives outcome from one or more hardware units and routes it to one or more decoder modules. The decoder modules solve a decoding problem specified by the decoding coordinator. They receive outcome data from the measurement outcome data router as well as any updated checks from the boundary condition data store. After completing their task, they store boundary condition data in the boundary condition data store, and report partial membrane data to the decoding coordinator.
}
\label{fig:sys_overview}
\end{figure*}

For a device-level implementation of modular decoding, the classical data from measurements, checks, partial membranes (i.e., decoder output) and logical measurement outcomes need to be relayed to several different processing units. In Fig.~\ref{fig:sys_overview} we present a system-level schematic for this data flow.

\end{document}